\documentclass[11pt,
               abstract=on
]{scrartcl}
\usepackage[a4paper, left=25mm, right=25mm, top=20mm, bottom=25mm]{geometry}
\usepackage[utf8]{inputenc}
\usepackage[ruled]{algorithm2e}
\SetKwFor{Until}{until}{do}{end}

\usepackage{amsmath}
\usepackage{amssymb}
\usepackage{bbm}
\usepackage{booktabs}
\usepackage[font=small,labelfont=bf,tableposition=top]{caption}
\usepackage{csquotes}
\usepackage{graphicx}
\usepackage{hhline}
\usepackage{xcolor}
\definecolor{dodgerblue}{RGB}{30,144,255}
\definecolor{steelblue}{RGB}{70,130,180}
\definecolor{darkred}{RGB}{178,34,34}
\definecolor{forestgreen}{RGB}{34,139,34}

\usepackage{hyperref}
\hypersetup{
    colorlinks=true,
    citecolor=blue, 
    citecolor=steelblue,
    linkcolor=forestgreen
    }
\usepackage{multirow}
\usepackage{subcaption}
\usepackage{tabularx}

\usepackage[scaled=0.9]{roboto}
\allowdisplaybreaks

\newcolumntype{Y}{>{\centering\arraybackslash}X}

\usepackage{authblk}
\setkomafont{author}{\scshape}

\subject{\footnotesize Preprint}

\title{Black-box Bayesian inference for\\ economic agent-based models}
\author[1,2]{\textsc{Joel Dyer}}
\author[3]{\textsc{Patrick Cannon}}
\author[1,2,4]{\textsc{J. Doyne Farmer}}
\author[3,5]{\textsc{Sebastian Schmon}}
\affil[1]{Institute for New Economic Thinking, Oxford}
\affil[2]{Mathematical Institute, University of Oxford}
\affil[3]{Improbable}
\affil[4]{Santa Fe Institute}
\affil[5]{Durham University}
\date{\small\today}

\newcommand{\bu}{\mathbf{u}}
\newcommand{\bx}{\mathbf{x}}
\newcommand{\bX}{\mathbf{X}}
\newcommand{\bU}{\mathbf{U}}
\newcommand{\by}{\mathbf{y}}
\newcommand{\bs}{\mathbf{s}}
\newcommand{\bth}{\boldsymbol{\theta}}
\newcommand{\bTh}{\boldsymbol{\Theta}}
\newcommand{\km}{\textsc{kde}}
\newcommand{\kms}{\km\ }
\newcommand{\mmd}{\textsc{mmd}}
\newcommand{\mmds}{\mmd\ }
\newcommand{\wass}{\textsc{Wasserstein}}
\newcommand{\wasss}{\wass\ }

\usepackage[acronym,smallcaps,nowarn,section,nogroupskip,nonumberlist,shortcuts]{glossaries}
\glsdisablehyper
\newacronym{abc}{abc}{approximate Bayesian computation}
\newacronym{lfi}{lfi}{likelihood-free inference}
\newacronym{sbi}{sbi}{simulation-based inference}
\newacronym{mcmc}{mcmc}{Markov chain Monte Carlo}
\newacronym{mh}{mh}{Metropolis-Hastings}
\newacronym{pmcmc}{pmcmc}{particle Markov chain Monte Carlo}
\newacronym{kde}{kde}{kernel density estimation}
\newacronym{npe}{npe}{neural posterior estimation}
\newacronym{snpe}{snpe}{\emph{sequential} neural posterior estimation}
\newacronym{smd}{smd}{simulated minimum distance}
\newacronym{nre}{nre}{neural ratio estimation}
\newacronym{snre}{snre}{\emph{sequential} neural ratio estimation}
\newacronym{mvgbm}{mvgbm}{multivariate geometric Brownian motion}
\newacronym{rkhs}{rkhs}{reproducing kernel Hilbert space}
\newacronym{sbc}{sbc}{simulation-based calibration}
\newacronym{sir}{sir}{sampling-importance-resampling}
\newacronym{msm}{msm}{Method of Simulated Moments}
\newacronym{ii}{ii}{Indirect Inference}
\newacronym{smc}{smc}{sequential Monte Carlo}
\newglossaryentry{abm}
{
  name={\textsc{abm}},
  description={agent-based model},
  first={\glsentrydesc{abm} (\glsentrytext{abm})},
  plural={\textsc{abm}s},
  descriptionplural={agent-based models},
  firstplural={\glsentrydescplural{abm} (\glsentryplural{abm})}
} 
\newglossaryentry{gru}
{
  name={\textsc{gru}},
  description={gated recurrent unit},
  first={\glsentrydesc{gru} (\glsentrytext{gru})},
  plural={\textsc{gru}s},
  descriptionplural={gated recurrent units},
  firstplural={\glsentrydescplural{gru} (\glsentryplural{gru})}
} 

\usepackage{xcolor}

\usepackage[english]{babel}

\usepackage[round]{natbib}
\begin{document}

\maketitle

\begin{abstract}
\noindent
    Simulation models, in particular agent-based models, are gaining popularity in economics. The considerable flexibility they offer, as well as their capacity to reproduce a variety of empirically observed behaviors of complex systems, give them broad appeal, and the increasing availability of cheap computing power has made their use feasible. Yet a widespread adoption in real-world modelling and decision-making scenarios has been hindered by the difficulty of performing parameter estimation for such models. 
    In general, simulation models lack a tractable likelihood function, which precludes a straightforward application of standard statistical inference techniques. A number of recent works \citep{grazzini2017bayesian, PLATT2020103859, platt2021bayesian} have sought to address this problem through the application of \emph{likelihood-free} inference techniques, in which parameter estimates are determined by performing some form of comparison between the observed data and simulation output. However, these approaches are (a) founded on restrictive assumptions, and/or (b) typically require many hundreds of thousands of simulations. These qualities make them unsuitable for large-scale simulations in economics and can cast doubt on the validity of these inference methods in such scenarios.
    In this paper, we investigate the efficacy of two classes of simulation-efficient black-box approximate Bayesian inference methods that have recently drawn significant attention within the probabilistic machine learning community: neural posterior estimation and neural density ratio estimation. 
    We present a number of benchmarking experiments in which we demonstrate that neural network based black-box methods provide state of the art parameter inference for economic simulation models, and crucially are compatible with generic multivariate time-series data.
    In addition, we suggest appropriate assessment criteria for use in future benchmarking of approximate Bayesian inference procedures for economic simulation models.
\end{abstract}

\section{Introduction
}

Simulation models are increasingly employed across the sciences. 
Their primary advantage is the flexibility they afford modellers: models may be specified mechanistically and at the microscopic level without concern for analytic tractability of the full system behaviour. 
This flexibility is embraced in the \emph{agent-based modelling} paradigm wherein individual agents can be described by realistic and heterogeneous decision and interaction rules. Examining the macroscopic properties of the system is then done simply through simulation, freeing the modeller from the burden of reasoning about the global behaviour of the system themselves and facilitating the study of radically more complex systems and emergent phenomena than is otherwise possible.

In general, \glspl{abm} take as input some parameter vector $\bth$, and return a (possibly multivariate) time-series $\bx$. 
Typically, \glspl{abm} are \emph{stochastic} simulators, with the property that running the simulator repeatedly with a fixed parameter $\bth$ will produce different outputs with each run. The probability (density) with which the simulator produces a particular output $\bx$, conditional on an input parameter vector $\bth$, is given by its likelihood function, written $p(\bx \mid \bth)$, which in some sense encapsulates the statistical properties of the \gls{abm}. We denote simulation of a dataset $\bx$ from the \gls{abm} as a draw from its probability density function, $\bx \sim p(\bx \mid \bth)$.

An essential prerequisite for the successful application of \glspl{abm} in real-world decision making scenarios is the ability to perform parameter estimation, that is, to tune $\bth$ such that the simulated data are good matches to observed data, which we denote with $\by$. Traditional statistical workhorses like maximum likelihood estimation or Bayesian inference rely on pointwise evaluations of the likelihood function. In practice, $p(\bx \mid \bth)$ cannot be obtained or evaluated in a reasonable time for arbitrary simulation models, and is thus only implicitly defined by the software specifying the behaviour of the simulator. This presents a barrier to the use of \glspl{abm} in real-world use.

To overcome this, a number of approaches to performing statistical inference have been developed in which exact density evaluations are replaced by evaluations of approximate densities or cost functions that are constructed using simulations from the model. Some of these \gls{sbi} methods have been explored within the \gls{abm} community. Perhaps the most prevalent of them is \gls{smd}, in which an estimate $\hat{\bth}$ is obtained by minimising some loss function $f(\by, \bth)$ between the observed data $\by$ and simulated data $\bx \sim p\left(\cdot \mid \bth\right)$ over some search space $\bTh$:
\begin{equation}
    \hat{\bth} = \arg\min_{\bth \in \bTh} f\left(\by, \bth \right).
\end{equation}
This general class of estimators includes the maximum likelihood estimator --  corresponding to choosing, for example, $f\left(\by, \bth\right) = -\log{p\left(\by \mid \bth\right)}$ or, where the likelihood is intractable, some estimate thereof \citep[see e.g.][]{diggle1984monte, kukacka2017estimation}.
It also includes the \gls{msm} \citep{franke2009applying} -- in which $f$ takes the form
\begin{equation}
    f(\by, \bth) = \left( g(\by) - \hat{g}_{\bth} \right)' 
    W 
    \left( g(\by) - \hat{g}_{\bth}\right),
\end{equation}
where $g(\by)$ denotes a set of moments derived from $\by$, $\hat{g}_{\bth}$ denotes the same set of moments derived from $R \geq 1$ simulations at $\bth$, $W$ is a suitably chosen weight matrix, and $'$ denotes the transpose.
As a final example, it includes \gls{ii} \citep{gourieroux1993indirect}, which follows a similar approach to \gls{msm} but replaces the moments with estimated parameters of a tractable auxiliary model. Some further loss functions have been proposed more recently in the context of \gls{abm} estimation \citep[see][for a recent review]{PLATT2020103859}.

A major drawback of \gls{smd} and optimisation-based 
approaches more generally is that only parameter point estimates are produced, while in many contexts it is desirable to obtain meaningful uncertainty quantification regarding appropriate values for $\bth$. 
Bayesian inference is an alternative inferential paradigm which naturally provides this meaningful notion of uncertainty, and has seen some recent but otherwise limited attention within the \gls{abm} community \citep[see e.g.][]{grazzini2017bayesian, lux2018estimation, PLATT2020103859, platt2021bayesian, lux2021bayesian}. Here, the primary object of interest is the parameter posterior distribution $p( \bth \mid \by )$, obtained via Bayes theorem: 
\begin{equation}\label{Eq:Bayes}
    p\left(\bth \mid \by\right) = \frac{p\left(\by \mid \bth\right) }{p\left(\by\right)}p\left(\bth\right).
\end{equation}
In Equation \eqref{Eq:Bayes}, $p\left(\bth\right)$ is referred to as the \emph{prior distribution} over parameters $\bth \in \bTh$, which encodes the experimenter's initial beliefs regarding appropriate values for $\bth$ before data is observed, while the simulation model itself appears via its associated data likelihood function $p\left(\by \mid \bth\right)$. Importantly, 
Bayesian inference procedures seek to accurately approximate the entire distribution $p\left(\bth \mid \by\right)$, rather than any single point estimate derived from this distribution. In this way, accurate and meaningful uncertainty estimates may also be obtained which correctly account for the experimenter's initial beliefs, as well as the evidence provided by the data $\by$.

A well-known investigation into Bayesian estimation methods for \glspl{abm}
is \citet{grazzini2017bayesian}, which explores various means of constructing a surrogate likelihood $\tilde{p}\left(\by \mid \bth\right)$ to account for the intractability of the true likelihood $p\left(\by \mid \bth\right)$. Two immediate drawbacks of the methods \citet{grazzini2017bayesian} discuss are that they (a) are difficult to reconcile with the fact that \glspl{abm} are frequently dynamical models generating time-series output, and (b) entail a huge computational burden, since each likelihood evaluation in 
the employed posterior sampling algorithm requires at least one simulation from the model, and it is typically necessary to make many hundreds of thousands of such evaluations. 
Recent work by \citet{platt2021bayesian} suggests estimating the model's transition density via a mixture density network.
While the approach offers some improvements through the use of a more flexible density estimator and the incorporation of some temporal dependencies, it suffers from similar drawbacks, assuming time-homogeneity and requiring a computationally expensive estimation step for every single sample from the approximate posterior distribution. In summary, there is a need for parameter inference methods that fulfil two main criteria: 

\begin{enumerate}
    \item they must be simulation-efficient in order to remain applicable to large-scale simulation models such as \glspl{abm}; 
    \item and they must be able to deal with non-homogenous/non-stationary temporal data, both simulated and observed.
\end{enumerate}

To this end, we seek with this paper to analyse the utility of two classes of 
parameter inference methods that have seen significant activity within the computational statistics and probabilistic machine learning literature in recent years: 
neural posterior estimation \citep{papamakarios2016fast, lueckmann2017flexible, Greenberg2019}, and neural density ratio estimation \citep{thomas2021lfire, pmlr-v119-hermans20a, Durkan2020}. 
Such methods have been employed successfully in a variety of applied domains, including high-energy physics \citep{brhemer2018constraining}, cosmology \citep{alsing2019fast}, and neuroscience \citep{gonccalves2020training}. In addition, as we will demonstrate below, they work flexibly with potentially multivariate time-series data, requiring minimal model assumptions, and typically generate more accurate parameter inferences with a significantly reduced simulation budget in comparison to common alternatives, such as those reported in \citet{grazzini2017bayesian}. 

We further seek to establish a higher standard of assessment criteria to be used in future benchmarking experiments for approximate Bayesian parameter inference for economic simulation models through the use of common integral probability metrics and inference validation procedures used widely in the statistics and machine learning communities. In Bayesian inference, the object of interest is the posterior density, whose entire shape is important since this captures degrees of beliefs regarding the unknown $\bth$. However, the literature on Bayesian estimation of \glspl{abm} has largely forgone attempts to examine the degree of overlap between the estimated and ground-truth posterior densities or the extent to which the inference procedure has learnt the dynamics of the data, 
often in favour of a simple comparison of certain point estimates of $\bth$ \citep{grazzini2017bayesian, platt2021bayesian, shiono2021estimation}. 
Since the objective of Bayesian inference is to derive a posterior probability density function which naturally quantifies uncertainty, it is important to adopt assessment criteria that quantify the degree of closeness between the estimated and target posterior densities in benchmarking experiments.

In summary, our contributions with this article are to provide:

\begin{enumerate}
    \item an overview of recent advances 
    in black-box simulation-based Bayesian inference for simulation models, and a motivation for their use in the specific case of agent-based models;
    \item a systematic benchmarking of these state-of-the-art methods against popular alternatives within the literature on parameter inference for agent-based economic simulation models;
    \item a discussion and demonstration of approaches to validating approximate Bayesian inference algorithms, and their feasibility for agent-based models under different simulation-based inference schemes.
\end{enumerate}

\section{Review of Bayesian inference methods for agent-based models}\label{sec:review}

Bayesian inference has seen relatively little attention from the economic \gls{abm} community as a means to parameter estimation. In this section, we summarise the key literature in this space. Importantly, we emphasise throughout that a common feature of these approaches is that the task of inference -- that is, of constructing the posterior distribution -- is inherently linked to the act of simulating from the \gls{abm} and building an additional approximate \emph{generative} model (sometimes also called an \emph{emulator}), in the sense that an approximation to the likelihood function is constructed. We will see in later chapters that this contrasts with a new generation of estimation methods in which inference is decoupled from the act of simulating and can be performed in a \emph{discriminative} manner, typically resulting in more efficient inferences.

\subsection{Approximate Bayesian computation}

Bayesian estimation of economic \glspl{abm} has gained popularity with the work of \citet{grazzini2017bayesian}, in which the authors discuss three approaches to approximate Bayesian parameter inference. In this section, we briefly outline the three approaches, and show that each method can seen as explicitly or implicitly approximating the unknown likelihood function associated with the \gls{abm} by
\begin{equation}\label{eq:abc_likelihood}
    \hat{p}(\by \mid \bth) \propto \int K_{\epsilon}\left(\by, \bx\right) p\left(\bx \mid \bth\right) {\mathrm{d} \bx},
\end{equation}
where $K_{\epsilon}$ is a (possibly unnormalised) kernel function, with parameter(s) $\epsilon$, providing a measure of \enquote{distance} between observed data $\by$ and simulated data $\bx$. This permits an approximation $\hat{p}\left(\bth \mid \by\right)$ of the posterior density as
\begin{equation}\label{eq:GrazzPost}
    \hat{p}\left(\bth \mid \by\right) \propto \int K_{\epsilon}\left(\by, \bx\right) p\left(\bx \mid \bth\right) p\left(\bth\right) {\mathrm{d} \bx}.
\end{equation}
Such approaches -- targeting \eqref{eq:GrazzPost} -- are comprehensively subsumed under \enquote{\gls{abc}} which, as noted by \citet{grazzini2017bayesian}, is a technique that has received significant attention within the computational statistics community over the last two decades as a means to constructing approximate posterior densities for models with intractable likelihood functions \citep[see e.g.][]{tavare1997inferring, pritchard1999population, beaumont2002approximate, sunnaaker2013approximate, gbiabc}. The approach has been successfully applied within a variety of fields, most prominently in ecology, epidemiology, and systems biology \citep{toni2009approximate, liepe2014framework, beaumont2010approximate, ju2021sequential}. 
While the approaches discussed by \citet{grazzini2017bayesian} can be cast in this general form, a judicious choice of a \emph{particular} kernel $K_\epsilon$ based on reasonable assumptions is crucial. In particular, different kernel choices are called for in different scenarios. Some possibilities are outlined below.

\subsubsection{Parametric density estimation}\label{sec:pde}

The simplest approach discussed by \cite{grazzini2017bayesian} involves assuming that the \gls{abm} has entered a statistical equilibrium, such that the observations $y_t \in \mathbb{R}^d$ in the time-series $\by := (y_1, y_2, \dots, y_T) \in \mathbb{R}^{
d\times T
}$ are fluctuations around some stationary value $m^{*}$:
\begin{equation}
    y_t = m^{*} + \varepsilon_t,
\end{equation}
where $(\varepsilon_t)$ are \emph{iid} noise terms with density $g_{\epsilon}$ and parameters $\epsilon$. Although more complex distributions are available, $g_{\epsilon}$ may for example be a zero-mean Gaussian distribution, and $\epsilon$ the elements of the covariance matrix. 
%
%

%
%
Under these assumptions, recalling that the elements of the time-series are assumed to be independent, the parameters $m^{*}$ and $\epsilon$ can be estimated by the means and covariances of the elements of the 
time-series to give $\hat{m}^*$ and $\hat{\epsilon}$. More precisely, first fixing $\bth$ and drawing a sample $\bx \sim p(\bx \mid \bth)$, one can calculate the estimates $\hat{m}^*(\bx)$ and $\hat{\epsilon}(\bx)$ using, for example, maximum likelihood estimation. Adopting as the measure of distance the probability of the true data being observed under this approximated process yields the choice of kernel
\begin{equation}
    K_{\epsilon}\left(\by, \bx\right) = \prod_{t=1}^{T} g_{\hat{\epsilon}\left(\bx\right)}
    \left(y_t - \hat{m}^* \left(\bx\right) \right).
\end{equation}
Finally by taking $R \geq 1$ \emph{iid} simulated datasets\footnote{While $R$ can be chosen to be any natural number, it is usually efficient to take $R=1$ when averages of estimators are concerned and run the Markov chain for an accordingly higher number of iterations \citep{bornn2017use, sherlock2017pseudo}.}
$\bx^{(r)} \sim p\left(\bx \mid \bth\right)$, $r = 1, \dots, R$, and calculating their associated estimators $\hat{m}^{*}\left(\bx^{(r)}\right)$ and $\hat{\epsilon}\left(\bx^{(r)}\right)$, the likelihood at $\bth$, i.e. Equation \eqref{eq:abc_likelihood}, is approximated with the Monte Carlo average
\begin{equation}\label{eq:pde}
    \hat{p}_{\bth}(\by) \approx \frac{1}{R} \sum_{r=1}^{R} \prod_{t=1}^{T} g_{\epsilon\left(\bx^{(r)}\right)}\left(y_t - m^{*}\left(\bx^{(r)}\right)\right),
\end{equation}
where we write $\hat{p}_{\bth}$ instead of $\hat{p}(\cdot \mid \bth)$.
Alternatively, the $R$ simulations may be pooled to generate single estimates $\hat{m}^{*}\left(\bx^{(1)}, \dots, \bx^{(R)}\right)$ and $\hat{\epsilon}\left(\bx^{(1)}, \dots, \bx^{(R)}\right)$ at $\bth$. This is closely related to \emph{synthetic likelihood} approaches \citep{wood2010statistical, price2018bayesian} which use similar Gaussian distributions, but usually rely on summary statistics.

In either case, the resulting approximate likelihood $\hat{p}_{\bth}$ from Equation \eqref{eq:pde} can then be used downstream for parameter estimation, either directly via e.g. maximum likelihood or through the corresponding approximate posterior, simulated e.g. through \gls{mcmc}. This approach suffers from clear limitations, however. Firstly, it treats the data points in the observed time-series as independent -- that is, as lacking a natural ordering -- which destroys important information when the observed $\by$ and simulated $\bx$ are time-series. Secondly, choosing an appropriate and sufficiently flexible family of parametric densities $g_{\epsilon}$ to construct the likelihood approximation is non-trivial, with poor choices leading to erroneous Bayesian inference. 

\subsubsection{Non-parametric density estimation}\label{sec:km}

An alternative approach, which partially addresses the second limitation described in Section \ref{sec:pde}, is to forgo the assumption of a parametric family of densities and instead use a non-parametric method for density estimation. \cite{grazzini2017bayesian} describe the use of \gls{kde} for this purpose. Here, the data points in the time-series are once again assumed to be independent and fluctuating about some stationary value $m^*$ as in the parametric approach described above. Then, an estimate of the likelihood function is obtained by applying \gls{kde} to $R \geq 1$ \emph{iid} simulations of length $S$, $\bx^{(r)} := (x^{(r)}_1, \dots, x^{(r)}_S) \sim p\left(\bx \mid \bth\right)$, $r = 1, \dots, R$, providing an unbiased estimate of the approximated likelihood function in Equation \eqref{eq:abc_likelihood} as
\begin{equation}\label{eq:npde}
    \hat{p}_{\bth}(\by) \approx \frac{1}{R} \sum_{r=1}^{R} K_{\epsilon}\left(\by, \bx^{(r)}\right) := \frac{1}{R} \sum_{r=1}^{R} \prod_{t=1}^{T} \hat{p}_{\epsilon}\left(y_t \mid \bth, \bx^{(r)}\right),
\end{equation}
where $\hat{p}_{\epsilon}\left(y_t \mid \bth, \bx^{(r)}\right)$ is the estimate of the conditional density $p\left(y_t \mid \bth, \bx^{(r)}\right)$ obtained via \gls{kde}:
\begin{equation}
    \hat{p}_{\epsilon}(y_t \mid \bth, \bx) = \frac{1}{S} \sum_{s=1}^{S} \kappa_{\epsilon}\left( y_t - x_s \right),
\end{equation}
for some probability kernel $\kappa_{\epsilon}$ with bandwidth parameter(s) $\epsilon$. In particular, the authors employ a Gaussian kernel with bandwidth chosen using Silverman's method \citep{silverman1986density}, such that
\begin{equation}
   \kappa_{\epsilon}\left( y_t - x_s \right) = \frac{1}{\epsilon} \kappa\left( \frac{\| y_t - x_s \|_2}{\epsilon} \right).
\end{equation}
Alternatively, as in Section \ref{sec:pde}, the $R$ simulations may be pooled and a single \gls{kde} model fit to the combined dataset to obtain $\hat{p}(\by \mid \bth)$. While these non-parametric approaches to density estimation are arguably more flexible than the assumption of a parametric family, they are well known to suffer from the \emph{curse of dimensionality}, limiting their applicability to low dimensional data only, even under the unrealistic assumption that the $y_t$ are assumed independent.

\subsubsection{Classical approximate Bayesian computation}

The authors discuss a third common approach to simulation-based Bayesian inference for intractable simulation models, namely classical \gls{abc}. A typical \gls{abc} algorithm proceeds by proposing a parameter value $\bth$ from some proposal distribution, 
for example the prior density, and determining whether to accept or reject $\bth$ on the basis of some meaningful notion of distance $d(\bx, \by)$ between a simulation $\bx \sim p(\bx \mid \bth)$ and the observed data $\by$. 
A common choice for the distance $d$ is the Euclidean distance between suitable summary statistics of the data; that is, $d(\bx, \by) := \lVert \bs(\bx) - \bs(\by) \rVert$ where $\bs$ is a vector of, for example, the mean, standard deviation, and lag-1 autocorrelation of the data. 
While many versions of \gls{abc} exist, the simplest -- Rejection \gls{abc} -- involves repeatedly proposing candidate parameter values $\bth \sim p(\bth)$ and rejecting in the event that $d\left(\bx, \by\right) > \epsilon$, where $\bx \sim p(\bx \mid \bth)$.
Rejection \gls{abc} can therefore be cast in the form of Equation \eqref{eq:GrazzPost} by taking
\begin{equation}
    K_{\epsilon}\left(\by, \bx\right) = \mathbbm{1}\left[d\left(\bx, \by\right) \leq \epsilon\right],
\end{equation}
providing an estimate 
of the (unnormalised) approximate likelihood function in Equation \eqref{eq:abc_likelihood} as
\begin{equation}\label{eq:abc}
    \hat{p}_{\bth}\left(\by\right) \propto \frac{1}{R} \sum_{r=1}^{R} \mathbbm{1}\left[d\left(\bx^{(r)}, \by\right) \leq \epsilon\right].
\end{equation}
As discussed by \cite{grazzini2017bayesian}, \gls{abc} enjoys a number of desirable properties, in particular (under mild regularity conditions) consistency with the true posterior distribution as $\epsilon \to 0$, provided no information loss is incurred by the choice of distance $d$. However, a major challenge to its use in practice is designing such a distance function: the success or failure of the approach often hinges on the engineering of summary statistics that are close to sufficient\footnote{A Bayesian sufficient statistic $\bs(\bx)$ for a model $p(\bx\mid\bth)$ is one for which $p(\bth \mid \bs(\bx)) = p(\bth \mid \bx)$. Low-dimensional sufficient statistics for arbitrary probabilistic models are generally unobtainable \citep{pitman1936sufficient, koopman1936sufficient, darmois1935sufficient}.}.
A number of works have investigated strategies for summary statistic selection in \gls{abc} \citep[e.g.][]{Fearnhead2012, blum2013comparative, wiqvist2019partially, Chen2020} and there has been some success in alternative, summary-free distances \citep[see e.g.][]{Park2016, Bernton2019, dyer2021approximate}.
It is however in general not possible to derive a low-dimensional sufficient statistic for arbitrary simulation models, and a poor choice of summary statistics can lead to an unacceptable loss of information from the original data, resulting in a posterior approximation of little value.

\subsubsection{Unifying the approaches
}

In Algorithm \ref{alg:abc}, we show how the approximate posterior \eqref{eq:GrazzPost}, with any of the three possibilities for $K_{\epsilon}$ discussed by \citet{grazzini2017bayesian}, may be targeted 
using a variant of the popular 
\gls{mh} algorithm \citep[e.g.][]{Hastings1970MonteCS}. If the likelihood estimate $\hat{p}_{\bth}(\by)$ is an unbiased, non-negative estimate of some desired target $p_{\bth}(\by)$, then Algorithm \ref{alg:abc} is an example of a pseudo-marginal \gls{mh} algorithm \citep[][]{andrieu2009pseudo}.
Note that while the sampling procedure in Algorithm \ref{alg:abc} exactly targets the posterior distributions described in \citet{grazzini2017bayesian}, many alternative posterior sampling algorithms exist \citep[see e.g.][]{beaumont2009adaptive}. 

We once again emphasise that a major disadvantage to applying the methods described by \citet{grazzini2017bayesian} to \glspl{abm} is that estimating the posterior involves simulating $R \geq 1$ times from the \gls{abm} at each proposed parameter value $(\bth_i)_{i=1,\ldots,n}$. Since \glspl{abm} are typically expensive to simulate, and $n$ is required to be many hundreds of thousands in order to accurately estimate the targeted posterior, such approaches can rapidly lead to a prohibitively large number of required simulations. 

\begin{algorithm}[t]
\SetAlgoLined
\textbf{Input:} Prior distribution $p(\cdot)$, observation $\by$, proposal distribution $q(\cdot \mid \bth)$, initial value $\bth_0$, number of iterations $n$\;
\KwResult{Empirical posterior $\sum_{i=1}^n \delta_{\bth_i}$}
  \For{$r=1,\dots,R$}{
    Simulate $\bx^{(r)} \sim p\left(\cdot \mid \bth_0\right)$\;
  }
 \For{$i=1,\dots,n$}{
  Sample $\bth \sim q\left(\cdot \mid \bth_{i-1}\right)$\;
  \For{$r=1,\dots,R$}{
    Simulate $\tilde{\bx}^{(r)} \sim p\left(\cdot \mid \bth\right)$\;
  }Evaluate $\hat{p}_{\bth}(\by)$ using $\lbrace{\tilde{\bx}^{(r)}\rbrace}_{r=1}^{R}$ according to either Equation \eqref{eq:pde}, \eqref{eq:npde}, or \eqref{eq:abc} as desired\;
  Set $\hat{p}_{\bth_i}(\by) = \hat{p}_{\bth}(\by)$, $\bth_i = \bth$ and $\lbrace{\bx^{(r)}\rbrace}_{r=1}^{R} = \lbrace{\tilde{\bx}^{(r)}\rbrace}_{r=1}^{R}$ with probability 
  \begin{equation*}
      \alpha = \min \Bigg\{ 1, \frac{\hat{p}_{\bth}(\by) p(\bth) q(\bth_{i-1} \mid \bth)}{\hat{p}_{\bth_{i-1}}(\by) p(\bth_{i-1}) q(\bth \mid \bth_{i-1})} \Bigg\},
  \end{equation*}
  otherwise set $\hat{p}_{\bth_i}(\by) = \hat{p}_{\bth_{i-1}}(\by)$ and $\bth_i = \bth_{i-1}$.
 }
\caption{
Metropolis-Hastings sampling scheme for \gls{abc}}
\label{alg:abc}
\end{algorithm}

\subsection{Latent variable models}

A class of models closely related to approximate Bayesian computation is the class of so-called \emph{latent variable models}. Here, the real data, $\by$, is assumed to be a noisy observation of an unobserved (latent) process $\bx$. 
Thus, in contrast to previous approaches, some discrepancy between $\by$ and $\bx$ is to be expected. 

Under the most basic scenario, the data obtained from the simulator are identically $x_t \sim p(x \mid \bth)$ and $p(\bx \mid \bth) = \prod_{t=1}^T p(x_t \mid \bth)$. The observed data is then assumed to have an error distribution, $g_\varepsilon$, say, such that the contribution made by one observation $y_t$ to the likelihood function is thus
\begin{equation}\label{eq:latent_likelihood}
    p(y_t \mid \bth) = \int g_\varepsilon(y_t \mid x_t)p(x_t \mid \bth) \mathrm{d}x_t
\end{equation}
and $p(\by \mid \bth) = \prod_{t=1}^T p(y_t \mid \bth)$.
We draw the attention of the read to the conceptual similarity to \eqref{eq:abc_likelihood}; however, \eqref{eq:latent_likelihood} is no longer an approximation but is the exact likelihood under the assumption of measurement error. (Similarly, \gls{abc} can be seen as exact if $K_\varepsilon$ describes the observational error, see e.g. \citealt{wilkinson2013approximate}.) A simple unbiased estimator of \eqref{eq:latent_likelihood} is
\begin{equation*}\label{eq:latent_estimator}
    \hat{p}(\by \mid \bth) = \prod_{t=1}^T \frac{1}{R} \sum_{r=1}^R g_\varepsilon\left(y_t \mid x_t^{(r)}\right),\qquad x_t^{(r)} \sim p\left(x \mid \bth\right)
\end{equation*}
for some $R\geq 1$\footnote{Note that opposed to earlier cases the choice $R=1$ is generally not optimal for products of unbiased estimators, see \citet{doucet2015efficient, sherlock2015efficiency, schmon2021large}.}, which can then also be used in Algorithm \ref{alg:abc}, for example, to target the associated approximate posterior. 
The independence assumption underlying the simulator data is, however, not realistic for \glspl{abm} and other time-series models, and incorporating the temporal aspect of data requires further structural assumptions, such as those of hidden Markov models.

\subsubsection{Hidden Markov Models and particle filters}

Particle filters, an instance of a broader class of \gls{smc} methods \citep[see][for a recent overview of \gls{smc} and their application to epidemiological \glspl{abm}]{ju2021sequential}
relax the independence assumption imposed on the output of the simulation model
and have previously been employed for Bayesian parameter inference for economic \glspl{abm} \citep{lux2018estimation, lux2021bayesian}.
However, such methods also involve making assumptions about the structure of the model -- specifically, the assumption of a state space structure with a particular error distribution. 
Furthermore, previous works have noted the significant computational burden required for \gls{smc} methods \citep{malleson2020simulating, lux2021bayesian}, which can arise due to the fact that a number of particles that grows exponentially with the model dimensions is often required to avoid failure modes such as particle collapse.

\subsection{Neural methods}

As an alternative to kernel density estimation, neural density estimators have previously been employed to perform Bayesian estimation of \glspl{abm}. 
\citet{platt2021bayesian} presents a method which diverges from the approaches of \citet{grazzini2017bayesian} in two main regards. Firstly, in contrast to the previous independence assumption, it assumes an autoregressive structure for the time-series model to better capture temporal correlations 
in the data.
In particular, the approach assumes a time-series model that is Markov of order $L$, that is, $p(x_t \mid x_{1:t-1}, \bth) = p(x_t \mid x_{t-L:t-1}, \bth)$, where $x_{1:t} = (x_1, \dots, x_t)$. In words, the distribution of the state at time $t$ depends only on the previous $L$ states (and $\bth$).
In light of this assumption, the full likelihood can be factorised as
\begin{equation}
    p_{\bth}(\by) = p(y_{1:L} \mid \bth) \prod_{t=L+1}^{T} p(y_{t} \mid y_{t-1}, \dots, y_{t-L}, \bth).
\end{equation}
Secondly, as a consequence of the autoregressive model structure, the method employs a \emph{conditional} density estimator $q_\phi$ to approximate the transition density function $p(x_t \mid x_{t-L:t-1}, \bth)$. The resulting likelihood approximation can be written 
\begin{equation}
    \hat{p}_{\bth}(\by) = \prod_{t=L+1}^{T} q_{\phi(\bth)}(y_{t} \mid y_{t-1}, \dots, y_{t-L}),
\end{equation}
where the contribution of the first $L$ terms is ignored. A mixture density network \citep{bishop1994mixture} assumes the role of the conditional density estimator $q_{\phi(\bth)}$ with parameters $\phi(\bth)$ trained for each $\bth$. Denoting the $L$ states preceding $y_t$ as $\by_{<t}$ for brevity, the particular form of the mixture density network is 
\begin{equation*}
    q_{\phi}(y_{t} \mid \by_{<t}) = \sum_{k=1}^K \alpha_k(\by_{<t}) \, \mathcal{N}(y_t \mid \mu_k(\by_{<t}), \Sigma_k(\by_{<t})) .
\end{equation*}
where $\alpha_k$ are the mixing coefficients, and $\mu_k, \Sigma_k$ are the mean and covariance matrix respectively, all of which are parameterised by neural networks and trained via maximum likelihood on $R$ \emph{iid} model samples $\bx^{(r)}\sim p(\bx \mid \bth)$, $r=1,\ldots, R$. The resulting likelihood estimate can then once again be used also with Algorithm \ref{alg:abc}, for example, to target the associated approximate posterior distribution. 

While the model structure described above accounts for the sequential nature of simulations generated by \glspl{abm} to some extent, the computational burden associated with the simulation of sufficient training data and the training of a new conditional density estimator at each $\bth$ renders it largely infeasible \citep{shiono2021estimation}. In other work, \citet{shiono2021estimation} examines a particular neural approach, with an invertible architecture and the ability to accommodate time-series data, for the specific purpose of estimating the posterior density for a New-Keyensian \gls{abm}. In contrast to both \citet{platt2021bayesian} and \citet{shiono2021estimation}, we take in this work a broad perspective on the subject of neural simulation-based inference for \glspl{abm}, by thoroughly contextualising the problem of Bayesian parameter estimation for economic \glspl{abm} and by incorporating both benchmarking tasks and an exploration of suitable assessment criteria.

\glsresetall

\section{A new generation of 
simulation-based inference methods}

In this section, we provide an overview of the two main 
\gls{sbi} methods we make use of in this paper: \gls{npe} and \gls{nre}, both of which -- as their names suggest -- employ neural networks\footnote{Note that neural networks are not essential here -- the procedures we describe require only flexible function approximators. We frame the discussion around neural networks primarily because they are a convenient choice of flexible function approximators in many settings.} to obtain an estimate of the posterior density $p(\bth \mid \by)$. We motivate the use of these methods for \glspl{abm} by framing them as \emph{discriminative} approaches to simulation-based parameter inference, and contrasting them with the \emph{generative} approaches described in Section \ref{sec:review} which seek to approximate the model's distribution (likelihood) using some form of probabilistic model. We then present the core elements of these methods, and further provide a discussion on how they may be leveraged flexibly and automatically to accommodate inference tasks involving high-dimensional and potentially multivariate time-series data, as is often the case in economic applications.

\subsection{Motivating black-box discriminative approaches to parameter inference}

The methods we endorse in this article -- \gls{npe} and \gls{nre} -- can be motivated by the fact that they are:

\begin{enumerate}
    \item simulation-efficient alternatives to more traditional approaches to \gls{sbi}, such as the \gls{abc} approaches described in Section \ref{sec:review};
    \item generic \gls{sbi} methods that treat the simulator as a black-box, thus making minimal assumptions about the structure and output of the simulation model;
    \item discriminative approaches to \gls{sbi}, in the sense that inference does not require a probabilistic model for the data $\bx$: instead of \emph{explaining} the mechanisms that create the data, we only need to \emph{distinguish} between realisations that arise from different parameter values. 
\end{enumerate}

\paragraph{Simulation efficiency} 
The algorithms described in Section \ref{sec:review} share a common pattern. For a fixed parameter $\bth$, \emph{iid} simulations $\bx^{(r)} \sim p(\bx \mid \bth)$, $r = 1,\ldots,R$, are sampled to produce a proxy likelihood $\hat{p}(\bx \mid \bth)$. 
Then, to generate $n$ approximate posterior samples via \gls{mcmc}, this procedure is performed at $n$ different values for $\bth$, where $n$ must typically be a large number -- often a few hundred thousand, if not orders or magnitude larger -- to ensure a low Monte Carlo error. 
Consequently, at least $nR$ simulations from the \gls{abm} are required in total for inference under these algorithms. Since \glspl{abm} can be very expensive to simulate, and the act of simulating from the \gls{abm} remains the primary bottleneck in Bayesian estimation procedures for \glspl{abm}, 
this simulation demand can quickly become infeasible.

In contrast, the methods for which we advocate here differ by eliminating the need to simulate when sampling from the posterior. Instead, they decouple the act of simulating from the task of constructing the posterior by employing powerful function approximators to learn -- in essence -- \emph{global} posterior density estimators on the basis of a limited number of simulations from across the parameter space of the \gls{abm}; that is, they learn functions $h : \mathcal{Y} \times \bTh \to \mathbb{R}$ which approximate the posterior density $p(\bth \mid \by)$ across the space of all possible values for $\bth$ and $\by$. This has the potential to significantly reduce the simulation burden associated with approximate Bayesian inference procedures, because the pointwise estimates of $p(\bth \mid \by)$ can borrow strength from, and share information between, one another; in contrast, the algorithms described in Section \ref{sec:review} consider each pointwise evaluation of $p(\bth \mid \by)$ as standalone density estimation tasks, such that no information can be shared between them. In this way, a large number of approximate posterior samples can be generated from an algorithm trained on what can in practice be a far smaller number of model samples than is required for the algorithms described in Section \ref{sec:review}.

Finally, the approaches we endorse here are equipped with a further efficiency benefit:  \emph{amortisation}. This phrase captures the fact that the global density estimators 
can be used to generate samples from an approximate posterior $\hat{p}(\bth \mid \by)$ for any data $\by$ without the need to further simulate from the \gls{abm}, which results from the fact that they are trained on the full space of possible values for $\bth$ \emph{and} $\by$. In contrast, applying the methods described in Section \ref{sec:review} to a new dataset $\by'$ would entail another $nR$ simulations from the \gls{abm}, multiplying the already high simulation costs.


\paragraph{Black-box inference methods} We saw in Sections \ref{sec:pde} and \ref{sec:km} that many inference approaches come with restrictive assumptions, for instance, stationarity of the simulated and observed time-series. 
In general, it is useful to dispense with these assumptions, since they can be difficult to verify and limit the applicability of these methods for arbitrary simulators. Instead, it can be preferable to employ black-box methods that make minimal assumptions about the model and are therefore generically applicable to arbitrary simulators. Doing so enables the modeller to concentrate resources on model design and implementation, rather than on developing bespoke inference algorithms for each new simulator. Furthermore, assumptions such as stationarity are known to be particularly poorly suited to certain economic simulation models such as \glspl{abm}, since these models are known and are even designed to produce non-equilibrium dynamics. Employing inference procedures that are able to handle such dynamics is therefore essential to the task of estimating generic \glspl{abm}.

\paragraph{Discriminative approaches to parameter inference} Discriminative tasks in machine learning are typically simpler than generative tasks, since generative tasks address larger problems than pure discrimination. This is intuitive: for example, it is typically easier for humans and computers alike to distinguish between images of cats and dogs than to generate them. Analogously, it is generally a simpler task to discriminate between complex time-series data than it is to generate such time-series.
The approaches described in Section \ref{sec:review} adopt a generative approach: they each -- either explicitly or implicitly -- seek to derive an approximation to the simulator's likelihood function using a probabilistic model, and thus seek to model the (probability density function of the) simulation output itself. Formally, this may be understood as learning the (stochastic) map $\bth \mapsto \bx$.
\Gls{npe} and \gls{nre}, in contrast, do not seek to model the simulation output: instead, they map from instances of the simulation output to certain target values which we will describe in more detail below. This can be formally thought of as learning the map $\bx \mapsto \bth$.
Such an approach to parameter estimation therefore embodies a fundamental departure from the approaches described in Section \ref{sec:review}. It has the potential to be particularly beneficial for \glspl{abm}, which are known to be able to produce complex, non-equilibrium dynamics that are especially difficult to model and generate.

\subsection{Neural posterior estimation}\label{sec:npe}

In this section, we provide an introduction to neural conditional density estimators and their use in posterior estimation tasks. The core idea underlying this class of simulation-based Bayesian inference techniques is to use such conditional density estimators to model the parameter posterior density directly. 
While various conditional density estimators can be used for this purpose, e.g. mixture density networks \citep{bishop1994mixture, papamakarios2016fast}, we focus here on the case of \textbf{normalising flows} \citep{tabak2010density, tabak2013family, rezende} due to their widespread use in various density estimation tasks, including in the areas of image generation \citep[e.g.][]{kingma2018glow} and physics \citep[e.g.][]{noe2019boltz}.

\subsubsection{Normalising flows}\label{sec:nflows}
\begin{figure}
    \centering
    \includegraphics[width=\textwidth]{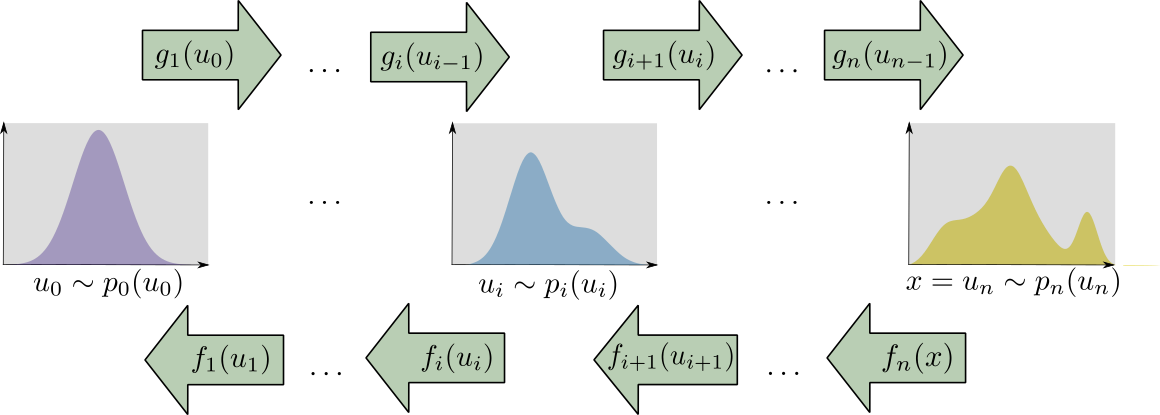}
    \caption{Schematic of a normalising flow. Sampling and density evaluation are performed via the processes illustrated at the top and bottom of the figure, respectively.}
    \label{fig:flow_sample_eval}
\end{figure}

Normalising flows involve transforming a simple base distribution into a more complicated one, often with the use of neural networks. Consider a random variable $\bU \sim p_{\bU}$, where $p_{\bU}$ is a probability distribution chosen to be a ``simple'' base distribution, in the sense that it is easy to both generate samples from $p_{\bU}$ and to evaluate $p_{\bU}(\bu)$ for any $\bu$.
Now consider the transformed random variable $\bX = g(\bU)$, where $g$ is a differentiable, invertible function with differentiable inverse $f := g^{-1}$. Then, by the change of variables formula, we have
\begin{align}\label{eq:prob_trans}
    p_{\bX}(\bx) = p_{\bU}\left( f(\bx) \right)\Big| \det{J_{f}(\bx)} \Big|
\end{align}
where $J_f$ is the Jacobian of $f$.
Sampling from $p_{\bX}$ then simply involves sampling a value $\bu$ from the base distribution $p_{\bU}$ and immediately obtaining $\bx = g(\bu)$. Evaluating $p_{\bX}(\bx)$ also then simply involves evaluation of the right-hand side of Equation \eqref{eq:prob_trans}. The above may be extended easily to a composition, or \emph{flow}, $g = g_{n} \circ g_{n-1} \circ \dots \circ g_{1}$ of differentiable and invertible functions $g_{i}$ with inverses $f_i$, for which we now have that
\begin{equation}
    p_{\bX}(\bx) = p_{\bU}\left(f_1\left( \dots f_{n-1}\left(f_{n}(\bx)\right) \dots \right)\right) \left\lvert{\prod_{i=1}^n\det{J_{f_i}(\bx)}}\right\rvert .
\end{equation}
Given samples from $p_{\bX}$, the question of how to choose 
a base distribution $p_{\bU}$ and a series of functions $g_1,\ldots,g_n$ such that the distribution of the samples is modelled accurately arises. 
The idea behind normalising flows is to \emph{learn} this sequence of transformations, that is, to have each $f_{i}$ be a flexible transformation $f_{\phi_i}$ with trainable parameters $\phi_i$. The resulting density estimator $q_{\phi}$ can be written
\begin{equation}
    q_{\phi}(\bx) = p_{\bU}\left(f_{\phi_1}\left( \dots f_{\phi_{n-1}}\left(f_{\phi_n}(\bx)\right) \dots \right)\right) \left\lvert{\prod_{i=1}^n\det{J_{f_{\phi_i}}(\bx)}}\right\rvert,\ \ \ \text{where}\ \ \phi = (\phi_n, \dots, \phi_1).
\end{equation}
The simple base distribution is then often taken to be a standard normal distribution\footnote{For this reason, the composition $f = f_{1} \circ \dots \circ f_{n-1} \circ f_{n}$ is often referred to as the \emph{normalising} flow, since it is a flow of transformations resulting (typically) in a Gaussian base distribution.} 
of the same dimension as the target distribution, and the flexible transformations $f_{\phi_i}$ are typically neural networks with a structure designed to guarantee the required invertibility and differentiability. The trainable parameters $\phi = \left\{ \phi_i \colon 1 \leq i \leq n \right\}$ are optimised by maximising the log-likelihood of the data $\bx^{(r)} \overset{iid}{\sim} p_{\bX}, r = 1, \ldots, R$:
\begin{equation}
    \hat{\phi} = \arg\max_{\phi} \sum_{r=1}^R \log{q_{\phi}\left(\bx^{(r)}\right)}.
\end{equation}
Normalizing flows are typically implemented in machine learning libraries that support automatic differentiation such as \texttt{pytorch} \citep{paszke2019pytorch} or \texttt{TensorFlow} \citet{abadi2016tensorflow} allowing the effective use of gradient-based optimisation techniques, such as Adam \citep[][]{kingma2014adam}. We provide a schematic of the sampling (flow, right-pointing arrows) and density evaluation (normalising flow, left-pointing arrows) processes in Figure \ref{fig:flow_sample_eval}, in which the simple, left-most distribution is morphed over successive steps 
into the more complex, right-most distribution.

\subsubsection{Normalising flows for neural posterior estimation}

Normalising flows may also be extended to the case of \emph{conditional} density estimation \citep[see e.g.][]{papamakarios2016fast, lueckmann2017flexible, papamakarios2019sequential, Greenberg2019}. In this way, we can use a normalising flow to estimate the posterior density $p(\bth \mid \bx)$ associated with a simulation model by following the same procedure as above and finding the neural network parameters as
\begin{equation}\label{eq:npe_loss}
    \hat{\phi} = \arg\max_{\phi\in\Phi} \sum_{r=1}^R \log{q_{\phi}\left(\bth^{(r)} \mid \bx^{(r)}\right)},
\end{equation}
where $(\bx^{(r)}, \bth^{(r)}) \overset{iid}{\sim} p(\bx, \bth), r=1, \ldots, R.$
Importantly, this framework allows us to learn a \emph{single} conditional density estimator for the posterior density function across data $\bx$, rather than having to undergo the expensive procedure of training a separate density estimator for each point evaluation of the posterior density as in e.g. \citet{platt2021bayesian}.

\subsection{Neural density ratio estimation}\label{sec:nre}

\glsreset{nre}

In this section, we describe an alternative common approach to \gls{sbi} termed \gls{nre}. The core idea underlying this method contrasts with \gls{npe} in that while \gls{npe} seeks to model the posterior density directly, \gls{nre} instead seeks to estimate the following ratio:
\begin{equation}
    \frac{p\left(\bx, \bth\right)}{p\left(\bx\right)p\left(\bth\right)} = \frac{p\left(\bx \mid  \bth\right)}{p\left(\bx\right)} = \frac{p\left(\bth \mid \bx\right)}{p\left(\bth\right)} =: r(\bx, \bth).
\end{equation}
This ratio is frequently referred to as the \textbf{likelihood-to-evidence ratio}. In the event that Bayesian inference is the goal, this then permits evaluation of the posterior density as
\begin{equation}
    p(\bth \mid \bx) = r(\bx, \bth) p(\bth).
\end{equation}
In the following, we discuss how this ratio may be efficiently estimated with well-calibrated probabilistic classifiers trained on simulated data-parameter pairs.

\subsubsection{Likelihood-to-evidence ratio estimation via probabilistic binary classification}
We outline the approach to approximate Bayesian inference via density ratio estimation as described by \citet{pmlr-v119-hermans20a}. Here, a probabilistic binary classifier is trained to distinguish between two sets of training examples:
\begin{enumerate}
    \item a set of ``genuine'' examples drawn from the joint distribution, $(\bx, \bth) \overset{iid}{\sim} p(\bx \mid \bth)p(\bth)$, which are assigned a class label $c=1$;
    \item a set of ``false'' examples drawn from the product of the marginals, $(\bx, \bth) \overset{iid}{\sim} p(\bx)p(\bth)$, with class label $c=0$.
\end{enumerate}
The difference between the two cases is that the $\bx$ are generated by the $\bth$ to which they are paired in the former set of examples, while in the latter set of examples the $\bx$ bear no relation to the $\bth$ they are paired with.

The function of a probabilistic binary classifier trained on such data is to model the class probability $d(\bx, \bth) := p(c = 1 \mid \bx, \bth) \in \left[0, 1\right]$; hard classification labels (i.e. decisions regarding the predicted value of $c$) are obtained when the continuous-valued $d(\bx, \bth)$ is combined with a decision rule, for example the rule that $c = 1$ should be predicted whenever $d(\bx, \bth) > 0.5$. One can show \citep[see Appendix B of][]{pmlr-v119-hermans20a} that the \emph{optimal} estimate of $d(\bx, \bth)$ is the value
\begin{equation}\label{eq:decision}
    d^{*}(\bx, \bth) := \frac{p(\bx \mid \bth)p(\bth)}{p(\bx \mid \bth)p(\bth) + p(\bx)p(\bth)}.
\end{equation}
Thus, a good probabilistic classifier trained to distinguish between the two possible types of pairs $(\bx, \bth)$ will learn a good estimate $\hat{d}(\bx, \bth)$ of this ratio.
Such a probabilistic classifier will allow us to evaluate the posterior density in this way: by noticing that one can rearrange Equation \eqref{eq:decision} as
\begin{equation}
    p(\bth \mid \bx) = \frac{d^{*}(\bx, \bth)}{1 - d^{*}(\bx, \bth)} p(\bth) =: r^{*}(\bx, \bth)p(\bth),
\end{equation}
where $r^{*}(\bx, \bth)$ is the corresponding estimate of the likelihood-to-evidence ratio $p(\bx \mid \bth)/p(\bx)$. In practice, of course, only an approximation $\hat{d}(\bx, \bth)$ will be obtained to the ratio in Equation \eqref{eq:decision}, yielding a correspondingly imperfect estimate $\hat{r}(\bx, \bth)$ of the likelihood-to-evidence ratio. Nonetheless, it is known that training expressive, high-capacity neural network classifiers via the cross-entropy loss
\begin{equation*}
    \ell\left(\left\{c^{(i)}, \bx^{(i)}, \bth^{(i)}\right\}_{i=1}^{N}\right) = -\frac{1}{N}\sum_{i=1}^{N} \left[ c^{(i)}\log{\hat{d}\left(\bx^{(i)}, \bth^{(i)}\right)} + \left(1 - c^{(i)}\right)\log{\left(1 - \hat{d}\left(\bx^{(i)}, \bth^{(i)}\right)\right)}\right]
\end{equation*}
can yield classifiers with good probability estimates, and thus good estimates of $r^{*}(\bx, \bth)$.

\subsubsection{A generalisation to multi-class classification}\label{sec:multiclassification}

In more recent work, \citet{Durkan2020} demonstrate that the above approach to density ratio estimation can be generalised to the problem of training a probabilistic classifier to 
identify the \emph{correct} $(\bx, \bth^{(i)})$ pair from a batch $\lbrace{(\bx, \bth^{(b)})}\rbrace_{b=1}^{B}$ of $B$ pairs that otherwise contain only ``incorrect'' pairs, where ``correct'' and ``incorrect'', respectively, correspond to having been drawn from the joint distribution $p(\bx, \bth)$ and from the product of the marginal distributions $p(\bx)p(\bth)$. In this case, the optimal estimate of the correct class probability is
\begin{align}
    p\left(c = i \mid \bx, \lbrace{\bth^{(b)}}\rbrace_{b=1}^{B}\right) &= \frac{p(\bth^{(i)} \mid \bx)\prod_{b\neq i}p(\bth^{(b)})}{\sum_{b=1}^{B} p(\bth^{(b)} \mid \bx)\prod_{b'\neq b}p(\bth^{(b')})}\\\label{eq:nreb_class}
    &= \frac{p(\bth^{(i)} \mid \bx)/p(\bth^{(i)})}{\sum_{b=1}^{B} p(\bth^{(b)} \mid \bx)/p(\bth^{(b)})}.
\end{align}
Thus,  by comparison with Equation \eqref{eq:nreb_class}, training a neural network $f_{\phi}$ on the loss function
\begin{equation}
    \ell(\phi) = -\log\frac{\exp{f_{\phi}(\bx, \bth^{(i)})}}{\sum_{b=1}^{B} \exp{f_{\phi}(\bx, \bth^{(b)})}}
\end{equation}
for each $(\bx, \bth) \sim p(\bx \mid \bth)p(\bth)$ will induce $f_{\phi}(\bx, \bth)$ to learn the value $f_{\phi}(\bx, \bth) = \log(p(\bth \mid \bx)/p(\bth))$, thus recovering an estimate of the desired density ratio. This can be extended to a batch of $R$ ``correct'' data-parameter pairs as
\begin{equation}\label{eq:nre_loss}
    \mathcal{L}(\phi) = -\frac{1}{R} \sum_{r=1}^{R} \log\frac{\exp{f_{\phi}(\bx^{(r)}, \bth^{(r)})}}{\sum_{b_r=1}^{B_r} \exp{f_{\phi}(\bx^{(r)}, \bth^{(b_r)})}},
\end{equation}
where the terms in the denominator are labelled with $b_r$ to account for the possibility that the contrasting (``incorrect'') set of parameters may be different for different ``correct'' pairs $(\bx^{(r)}, \bth^{(r)})$. The authors further demonstrate that this can yield more accurate density ratio estimators. For this reason, we adopt this approach throughout this article.

\subsection{Sampling from the neural posterior}

After successful training, the posterior density estimator $q_{\hat\phi}(\bth \mid \bx)$ obtained from \gls{npe} is a parametric approximation of the true posterior distribution, which can then be used to generate \emph{iid} samples $\bth \sim q_{\hat\phi}(\bth \mid \bx)$ or to evaluate the posterior density on a pointwise basis in the ways described in Section \ref{sec:nflows}. 
While ratio estimation similarly permits a pointwise evaluation of the posterior distribution as $p(\bth)\exp{(f_{\hat\phi}(\bx, \bth))}$, it requires a further inference step using, for example, Metropolis--Hastings to generate samples from the same posterior. However, the upfront training of the ratio estimator eliminates the need for 
expensive sampling from the \gls{abm}, 
reducing the run-time significantly in comparison to alternative approaches such as those described in Section \ref{sec:review}.

\subsection{Round-based training}

The procedures described in Sections \ref{sec:npe} and \ref{sec:nre} are framed as single density (ratio) estimation tasks. 
This means that a single dataset of $R$ data-parameter pairs $ (\bx^{(r)}, \bth^{(r)}), r = 1, \ldots, R$ is created in the following way:
\begin{align*}
    \bth^{(r)} &\sim p(\bth), & \text{sample from the prior;} \\
    \bx^{(r)}  &\sim p(\bx \mid \bth^{(r)}), & \text{sample from the model using the draws from the prior.}
\end{align*}
Subsequently, \gls{nre} and \gls{npe} algorithms are trained on the \emph{whole dataset} to find 
either the likelihood-to-evidence ratio or the posterior directly following, respectively, \eqref{eq:npe_loss} or \eqref{eq:nre_loss}.
Density estimators trained in this way are referred to as \textit{amortised} density estimators, which reflects the fact that they may be used to construct posterior distributions for any data $\bx$ without further simulation from the \gls{abm} or the need to train multiple density estimators. 

The disadvantage of this one-stage approach is that all training simulations from the \gls{abm} are generated by parameters drawn from the prior distribution. In many cases, however, interest lies only in the posterior for the \emph{particular observation} $\by$, which can be concentrated on specific subregions of the entire space covered by the prior density.
It can then be preferable to narrow down the search space of good candidate values for the parameter $\bth$ to subregions of the parameter space that could most plausibly have generated $\by$. 
By doing so, the density (ratio) estimator will be presented with a less varied 
range of dynamics between which it must learn to distinguish, allowing it to develop a more refined approximation of the density (ratio) in regions of high posterior density. This can facilitate more rapid learning and potentially reduce the number of training examples that must be simulated by the \gls{abm}.

Significant effort has thus been extended towards the constructions of \enquote{round}-based approaches to training density (ratio) estimators for \gls{sbi} \citep[see e.g.][]{papamakarios2019sequential, Greenberg2019, pmlr-v119-hermans20a, Durkan2020}.
The idea is to split the total budget of $R$ simulations into subsets of size $N_1, N_2, \ldots$ such that $\sum_i N_i = R$. In the first step, $N_1$ data points are created as before and a posterior approximation $q_{\hat\phi}\big(\bth \mid \bx\big)$ is constructed. 
Subsequently, in round $i \geq 2$, we create new data $(\bx^{(r)}, \bth^{(r)}), r = 1, \ldots, N_i$ using
\begin{align*}
    \bth^{(r)} &\sim q^{(i-1)}_{\hat\phi}\big(\bth \mid \bx\big), & \text{sample from round-$(i-1)$ posterior;} \\
    \bx^{(r)}  &\sim p(\bx \mid \bth^{(r)}), & \text{sample from the model using the draws from the round-$(i-1)$ posterior.}
\end{align*}
The posterior $q_{\phi}\big(\bth \mid \bx\big)$ may now be retrained using (a combination of the old and) the new samples, and the process can be repeated for as many rounds as necessary. An identical process can be followed for density ratio estimation, with the exception that parameters in round $i$ are drawn using the likelihood-to-evidence ratio obtained from round $i-1$ and, for example, Metropolis--Hastings.

While the round-based approach is appealing, it does not come without additional challenges. In particular, the joint distribution of the example pairs $(\bx^{(r)}, \bth^{(r)})$ is no longer $p(\bx, \bth)$ for data sampled after the first round. 
In practice, this needs to be accounted for during training by the use of additional weighing factors, such as those appearing in Equation \eqref{eq:snpe_loss} (cf. Equation \eqref{eq:npe_loss}).
We refer the interested reader to the following papers for further details on this matter: \citet[][]{papamakarios2016fast, lueckmann2017flexible, Greenberg2019}.

Training schemes for \gls{npe} and \gls{nre} which make use of this round-based training approach are referred to as \gls{snpe} and \gls{snre}, respectively, in which the prefix ``sequential'' reflects the round-based training design of the now non-amortised density (ratio) estimator.

\subsection{Summarising the simulation output}\label{sec:ss}


As described above, \gls{npe} and \gls{nre} 
take as input either $\bx$ alone or $(\bx, \bth)$ pairs. Given that $\bx$ is a (possibly multivariate) time-series for many \glspl{abm} and is thus a high-dimensional object, it can be beneficial to incorporate useful inductive biases\footnote{Inductive biases are usually realised as constraints on the form of the network architecture in the case of neural networks. For example, when certain symmetries are known to be present in the data, this inductive bias may be incorporated with the use of (partially) exchangeable neural networks \citep{NIPS2017_f22e4747, wiqvist2019partially}. Imposing such constraints helps to restrict the space over which appropriate functions are searched for and facilitates effective learning and generalisation from training data.} 
into the network architecture that account for the sequential nature of $\bx$ and reduce its dimensionality. 
To this end, the density (ratio) estimator can be prefixed with a so-called embedding network with trainable parameters $\varphi$, whose function is to consume the original high-dimensional dataset $\bx$ and express this as a lower-dimensional summary statistic vector $\bs_{\varphi}(\bx)$. The parameters of the embedding network and of the density (ratio) estimator may then be learned concurrently using the same loss function, offering a means to automatically learn descriptive low-dimensional features of the raw input data during the \emph{same} posterior or density ratio estimation task in an end-to-end fashion.

This quality is a further attractive feature of \gls{npe} and \gls{nre}. Finding low-dimensional summary statistics of the data is necessary in many approaches to \gls{sbi}\footnote{This is the prototypical approach to \gls{abc}, for example, although methods that obviate the need to summarise high-dimensional data have been proposed and explored in recent years \citep[see e.g.][]{Park2016, Bernton2019, dyer2021approximate}.}, and the question of how to summarise data naturally arises in response to this. Popular solutions that have been explored to-date include: the experimenter themselves devising a collection of hand-crafted summary statistics that they believe adequately describes the data, which is an arduous task that can lead to a difficult-to-quantify loss of information; or learning summary statistics by constructing a separate learning task to the existing problem of density (ratio) estimation \citep[see e.g.][]{Fearnhead2012, Chen2020}, imposing an additional burden on the experimenter. \Gls{npe} and \gls{nre}, in contrast, offer a convenient approach to incorporating inductive biases and to learning summary statistics in an end-to-end fashion without the additional complications associated with alternative \gls{sbi} techniques.


\subsection{Training procedures for density (ratio) estimators}

To summarise and conclude this section, we follow \citet{Greenberg2019} and \citet{Durkan2020} and provide in Algorithms \ref{alg:training_snpe} and \ref{alg:training_snre} procedures for 
training neural posterior and density ratio estimators, respectively, for \gls{sbi} over multiple rounds. In both cases, we assume that the density (ratio) estimator includes any embedding network used to learn summary statistics concurrently with the density (ratio) estimate, and thus consider $\phi$ to contain the parameters of both the estimator and the embedding network.

\begin{algorithm}[t]
\SetAlgoLined
\textbf{Input:} prior distribution $p(\bth)$, simulator $p(\bx \mid \bth)$, observation $\by$, conditional density estimator $q_{\phi}$, number of rounds $M$, number of simulations per round $N$\;
\KwResult{Trained conditional density estimator}
 Set $\tilde{p}_{0}(\bth) = p(\bth)$, dataset $\mathcal{D} = \lbrace{\rbrace}$\;
 \For{$m=0,\dots,M-1$}{
  Sample $\bth^{(n)} \overset{iid}{\sim} \tilde{p}_{m}(\bth)$, $n=1, \dots, N$\;
  Simulate $\bx^{(n)} \sim p\left(\bx \mid \bth^{(n)}\right)$, $n=1, \dots, N$\;
  {Append the dataset:
  \begin{equation*}
      \mathcal{D} := \mathcal{D} \cup \bigcup_{n = 1}^{N} \Big\{\left(\bx^{(n)}, \bth^{(n)}\right)\Big\};
  \end{equation*}
  }\\
  \Until{convergence}{
        Evaluate the loss function
        \begin{equation}\label{eq:snpe_loss}
            \mathcal{L}(\phi) = - \sum_{(\bx, \bth) \in \mathcal{D}} 
            \log{ \left(q_{\phi}(\bth \mid \bx) \frac{\tilde{p}_{m}(\bth)}{p(\bth)} \frac{1}{Z(\bx, \phi)} \right) };
        \end{equation}\\
        Update trainable parameters $\phi$ on the basis of $\mathcal{L}(\phi)$
    }
  Set $\tilde{p}_{m+1}(\bth) := q_{\phi}\left(\by, \bth\right)$.
 }
\caption{Training 
\gls{snpe} \citep[see][Algorithm 1]{Greenberg2019}.}
\label{alg:training_snpe}
\end{algorithm}

\begin{algorithm}[t]
\SetAlgoLined
\textbf{Input:} prior distribution $p(\bth)$, simulator $p(\bx \mid \bth)$, observation $\by$, density ratio estimator $f_{\phi}$, number of rounds $M$, number of simulations per round $N$, minibatch size $B$, contrasting set size $K$\;
\KwResult{Trained density ratio estimator}
 Set $\tilde{p}_{0}(\bth) = p(\bth)$, dataset $\mathcal{D} = \lbrace{\rbrace}$\;
 \For{$m=0,\dots,M-1$}{
  Sample $\bth^{(n)} \overset{iid}{\sim} \tilde{p}_{m}(\bth)$, $n=1, \dots, N$\;
  Simulate $\bx^{(n)} \sim p\left(\bx \mid \bth^{(n)}\right)$, $n=1, \dots, N$\;
  {Append the dataset:
  \begin{equation*}
      \mathcal{D} := \mathcal{D} \cup \bigcup_{n = 1}^{N} \Big\{\left(\bx^{(n)}, \bth^{(n)}\right)\Big\};
  \end{equation*}
  }\\
  \Until{convergence}{
        Sample minibatch $\Big\{\left(\bx^{(b)}, \bth^{(b)}\right)\Big\}_{b=1}^{B}$ uniformly from $\mathcal{D}$\;
        For each minibatch pair, draw $0 < K < B$ parameters $\tilde{\bth}^{(k)}$, $k = 1, \dots, K$, from elsewhere in     the training data $\mathcal{D}$\;
        Evaluate the finite-sample multinomial logistic loss
        \begin{equation}
            \mathcal{L}(\phi) = - \frac{1}{B} \sum_{b=1}^{B} 
            \log{ \frac{ \exp{\left( f_{\phi}\left(\bx^{(b)}, \bth^{(b)}\right) \right)} }
            { \exp{\left( f_{\phi}\left(\bx^{(b)}, \bth^{(b)}\right) \right)} + \sum_{k=1}^{K} \exp{\left( f_{\phi}\left(\bx^{(b)}, \tilde{\bth}^{(k)}\right) \right)} } };
        \end{equation}\\
        Update trainable parameters $\phi$ on the basis of $\mathcal{L}(\phi)$
    }
  Set $\tilde{p}_{m+1}(\bth) \propto \exp{\left(f_{\phi}\left(\by, \bth\right)\right)}$.
 }
\caption{Training 
\gls{snre} \citep[see][Algorithm 1]{Durkan2020}.}
\label{alg:training_snre}
\end{algorithm}


\section{Experiments for tractable examples}\label{sec:tractable}

In this section, we present experiments in which we compare the ability of \gls{npe} and \gls{nre} to estimate parameter posterior distributions for economic simulation models with tractable likelihood functions against the non-parametric density estimation method described in \citet{grazzini2017bayesian}, which we term \kms and outline in Section \ref{sec:km}. We provide details of the neural network architecture and training hyperparameters in Appendix \ref{app:nn}.

\subsection{Performance metrics}

Each example presented in this section will be equipped with a tractable transition density, and therefore a tractable likelihood function. Such models are thus amenable to Bayesian inference via standard means, such as \gls{mcmc}, to obtain an approximate ground-truth posterior density\footnote{While the true posterior density is targeted with such a sampling scheme, the ground-truth obtained in this fashion remains an approximation due to the Monte Carlo error associated with the finite sample size.}.

To assess the accuracy of the estimated posteriors in this case, we compare the full ground-truth posterior density with the full posterior estimated with each of the implemented simulation-based approaches. This contrasts with previous studies of Bayesian parameter estimation for \glspl{abm}, in which point estimates alone are often used to assess the quality of the tested inference procedures. For example, \citet{platt2021bayesian} uses a prior-weighted Euclidean distance between the estimated posterior mean and generating parameters (that is, the $\bth$ that generated the pseudo-observation $\by$) as a performance metric.
However, such metrics provide a very limited and at times misleading view on the outcome of the inference process, because (a) \enquote{closesness} is not measured in the geometry of the target distribution, as visualised in \autoref{fig:distance}; 
(b) it is possible that an approximate posterior can produce a posterior mean close to the generating parameter but simultaneously under- or over-estimate the width of the distribution or otherwise yield poor uncertainty quantification, and (c) finite datasets are not guaranteed to yield posterior densities that concentrate on the generating parameter.

To compare the ground-truth and simulation-based posteriors, we compute two integral probability metrics which each correspond to a notion of \emph{dissimilarity} between the ground-truth and estimated posteriors:

\begin{figure}
    \centering
    \includegraphics[width=0.8\textwidth]{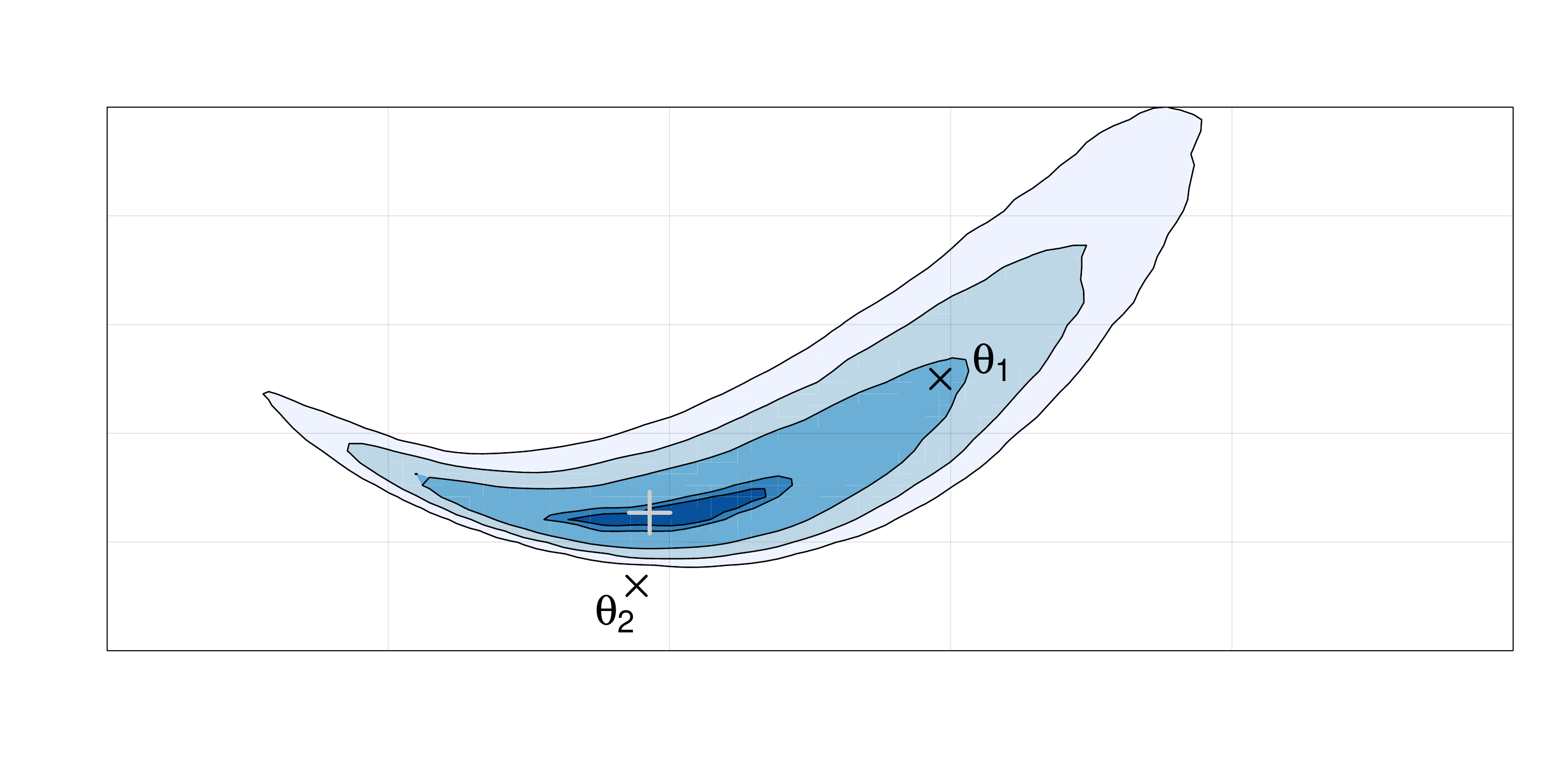}
    \caption{Visualisation: why standard Euclidean distances are misleading when gauging the performance of Bayesian inference algorithms. The parameter $\theta_2$ is ostensibly closer to the``true parameter" (grey) in this toy posterior than $\theta_1$. However, $\theta_1$ has higher posterior density, i.e. it is a more credible parameter given the observed data.}
    \label{fig:distance}
\end{figure}

\paragraph{Wasserstein distance}  

This is a distance measure derived from optimal transport theory \citep{wass} and used widely in various machine learning contexts, e.g. generative adversarial networks \citep{wgan}. For some distance $\rho_0$ on $\bTh$ and two probability measures $\mu$ and $\mu'$, the $p$-Wasserstein metric between two sets of samples $\lbrace{\bth_i}\rbrace_{i=1}^{n} \overset{iid}{\sim} \mu$ and $\lbrace{\bth'_i}\rbrace_{i=1}^{m} \overset{iid}{\sim} \mu'$ is computed as
\begin{equation}\label{eq:wass}
    \mathcal{W}_p\left(\lbrace{\bth_i}\rbrace_{i=1}^{n}, \lbrace{\bth'_j}\rbrace_{j=1}^{m}\right)^p = \inf_{\gamma \in \Gamma_{n,m}} \sum_{i =1}^{n} \sum_{j=1}^{m} \rho_0(\bth_i, \bth'_j)^p \gamma_{ij}
\end{equation}
where $\Gamma_{n,m}$ is the set of $n\times m$ matrices with non-negative entries, columns summing to $m^{-1}$, and rows summing to $n^{-1}$. Throughout, we use the Euclidean distance for $\rho_0$.

\paragraph{Maximum mean discrepancy (MMD)} This metric is once again a metric on probability distributions that draws from the theory of reproducing kernel Hilbert spaces \citep{gretton2006kernel, mmd} and is used widely within the machine learning and simulation-based inference community to assess the dissimilarity between probability distributions \citep{papamakarios2019sequential, lueckmann2021benchmarking}. Here, under a suitable choice of kernel\footnote{That is: a symmetric, positive semi-definite function on $\bTh$.} $k$ chosen by the experimenter, the discrepancy between two probability distributions $P$ and $Q$ is taken to be
\begin{equation}
    \textsc{mmd}(P, Q) = \Big\|\mathbb{E}_{\bth\sim P}\left[k(\bth, \cdot)\right] - \mathbb{E}_{\bth'\sim Q}\left[k(\bth', \cdot)\right] \Big\|_{\mathcal{H}}^2,
    \end{equation}
where $\mathcal{H}$ is the \gls{rkhs} associated with $k$ and $\mathbb{E}_{\bth\sim P}[k(\bth, \cdot)]$ is the so-called \emph{mean embedding} of $P$ via kernel $k$. When only samples $\lbrace{\bth_i}\rbrace_{i=1}^{n} \overset{iid}{\sim} P$ and $\lbrace{\bth'_i}\rbrace_{i=1}^{m} \overset{iid}{\sim} Q$ are available, an unbiased estimator of this metric can be computed as
\begin{equation}
    \widehat{\textsc{mmd}}\left(P, Q\right) = \frac{1}{m(m-1)} \sum_{\substack{i,j\\ j \neq i}} k(\bth'_i, \bth'_j) + \frac{1}{n(n-1)} \sum_{\substack{i,j\\ i \neq j}} k(\bth_i, \bth_j) - \frac{2}{n m} \sum_{i,j} k(\bth_i, \bth'_j).
\end{equation}
Throughout, we use a Gaussian kernel as $k$,
\begin{equation}
    k(\bth, \bth') = \exp\left(-\frac{\| \bth - \bth' \|_2^2}{2\sigma^2}\right),    
\end{equation}
where the scale parameter $\sigma^2 = \text{median}\lbrace{\| \tilde{\bth}_i - \tilde{\bth}_j \|^2_2\rbrace}$, following \citet{Briol2019}, and $\lbrace{\tilde{\bth}_i}\rbrace_{i=1}^{n}$ are the samples drawn from the ground-truth posteriors. 

\subsection{Example 1: Brock and Hommes (1998)}\label{sec:BH}

We consider a variant of the model proposed by \citep{BROCK19981235}, which has previously been used in \gls{abm} calibration experiments \citep{PLATT2020103859}. The model dynamics can be expressed as the following system of coupled equations:
\begin{align}
    x_{t+1} &= \frac{1}{R}\left[\sum_{h=1}^{H} n_{h, t+1}\left(g_h x_t + b_h\right) + \epsilon_{t+1}\right],\ \epsilon_{t} \sim \mathcal{N}(0, \sigma^{2}),\\
    n_{h, t+1} &= \frac{\exp{\left(\beta U_{h,t}\right)}}{\sum_{h' = 1}^{H} \exp{\left(\beta U_{h',t}\right)}},\\
    U_{h,t} &= \left(x_t - R x_{t-1}\right)\left(g_h x_{t-2} + b_h - R x_{t-1}\right),
\end{align}

\noindent where $R, \beta, \sigma$ are parameters. We follow \citet{PLATT2020103859} and assume that $H=4, R = 1.0, \sigma = 0.04, g_1 = b_1 = b_4 = 0$ and $g_4 = 1.01$. $\beta$ will be $120$ or $10$, depending on the experiment, and we note below which value is used. We consider the task of estimating the posterior $p\left(\bth \mid \by\right)$, where $\bth = \left(g_2, b_2, g_3, b_3\right)$, $\by := (y_1, y_2, \dots, y_T) \sim p(\bx \mid \bth^{*})$ is the pseudo-observation, $T=100$, and $\bth^{*}$ is the parameter setting used to generate $\by$. We consider two experimental setups which we describe further in the following two subsections. 
For each experiment, we note that by rewriting the above system of equations, we are able to find the transition density for observation $y_{t+1}$ as
\begin{equation}
    p(y_{t+1} \mid y_{1:t}, \bth) = \mathcal{N}\left(f(y_{t-2:t}, \bth), \frac{\sigma^2}{R^2}\right)
\end{equation}
where
\begin{equation}
    f\left(y_{t-2:t}, \bth\right) = \frac{1}{R}\sum_{h=1}^{H} \frac{\exp{\left[\beta \left(y_t - R y_{t-1}\right)\left(g_h y_{t-2} + b_h - R y_{t-1}\right)\right]}}{\sum_{h' = 1}^{H} \exp{\left[\beta \left(y_t - R y_{t-1}\right)\left(g_{h'} y_{t-2} + b_{h'} - R y_{t-1}\right)\right]}}\left(g_h y_t + b_h\right).
\end{equation}
In this way, we are able to obtain approximate ground truth posteriors with standard \gls{mcmc} techniques such as \gls{mh}.

For both \gls{npe} and \gls{nre}, we use a round-based training approach i.e. \gls{snpe} and \gls{snre}: we train over 10 rounds and generate 1000 simulations in each round. For \km, we take $R = 1$ and sample from the posterior with \gls{mh} (see Appendix \ref{app:mh} for further details).

\paragraph{Summarising data} As discussed in Section \ref{sec:ss}, there exists a number of approaches to representing high-dimensional time-series data $\bx$ as a low-dimensional vector $\bs(\bx)$ of summary statistics to facilitate \gls{sbi}. 
In the experiments for the Brock \& Hommes model, we demonstrate \gls{npe} and \gls{nre} using summary statistics obtained in two different ways: 

\begin{enumerate}
    \item by summarising them manually as the mean value, variance, maximum, minimum, median, 25th quantile, 75th quantile, and the autocorrelations of the $x_t$ to lags 1, 2, and 3\footnote{We note however that these are not the only choices available to the experimenter.}. We denote these summary statistics with $\tilde{\bs}$ and refer to them as \emph{naive} or \emph{hand-crafted} summary statistics;
    \item by learning them as part of the training process with an embedding network $\bs_{\varphi}$ with trainable parameters $\varphi$ (see Section \ref{sec:ss}). Through the experiments we present below, we will demonstrate that \gls{npe} and \gls{nre} can be used flexibly with various embedding networks,  
    and that the experimenter is free to choose from the plethora of candidate networks 
    that incorporate useful inductive biases for time-series data \citep[e.g.][]{Wong_2018, dyer2021deep, kidger2020neural}. Below, we refer to summary statistics obtained in this fashion as \emph{learned} summary statistics.
\end{enumerate}

\subsubsection{Parameter set 1}

We take $\beta = 120$ and consider the task of estimating the posterior density $p(\bth \mid \by)$, where $\bth^{*} := (g_2^{*}, b_2^{*}, g_3^{*}, b_3^{*}) = (0.9, 0.2, 0.9, -0.2)$. We use the following uniform priors: $g_2, b_2, g_3 \sim \mathcal{U}\left(0, 1\right)$, while $b_3 \sim \mathcal{U}(-1,0)$.

\begin{figure} 
  \begin{subfigure}[b]{0.48\linewidth}
    \centering
    \includegraphics[width=0.85\linewidth]{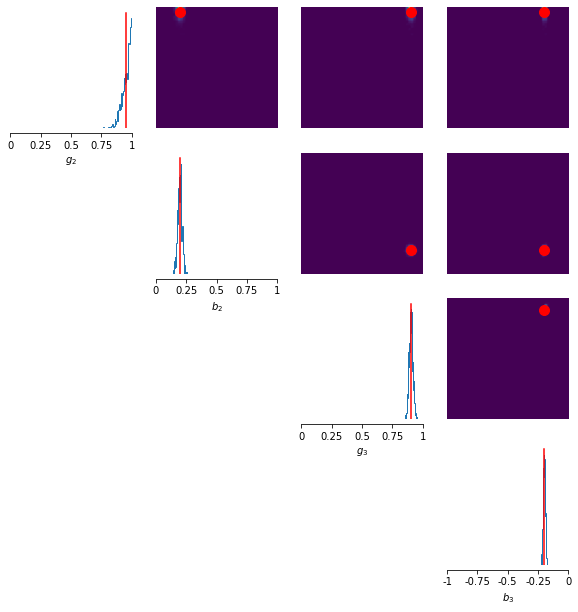} 
    \caption{True posterior} 
    \label{fig:BH1True} 
    \vspace{4ex}
  \end{subfigure}
  \hfill
  \begin{subfigure}[b]{0.48\linewidth}
    \centering
    \includegraphics[width=0.85\linewidth]{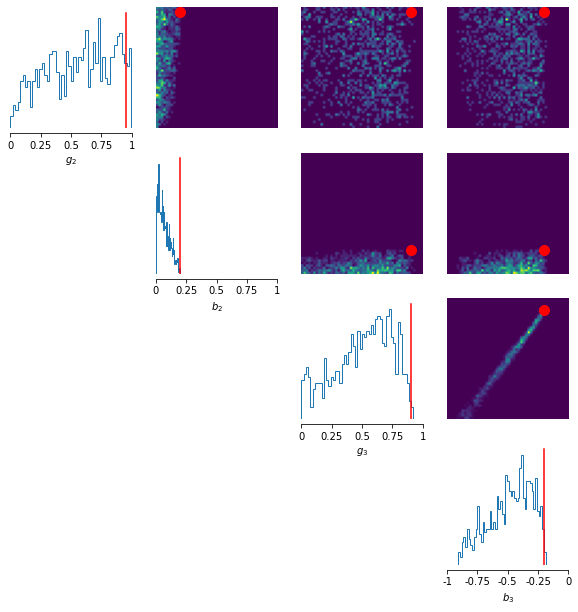}
    \caption{
             \textsc{kde} posterior, $1.5 \times 10^5$ simulations} 
    \label{fig:BH1G} 
    \vspace{4ex}
  \end{subfigure} 
  
  \vskip -0.1in
  \begin{subfigure}[b]{0.48\linewidth}
    \centering
    \includegraphics[width=0.85\linewidth]{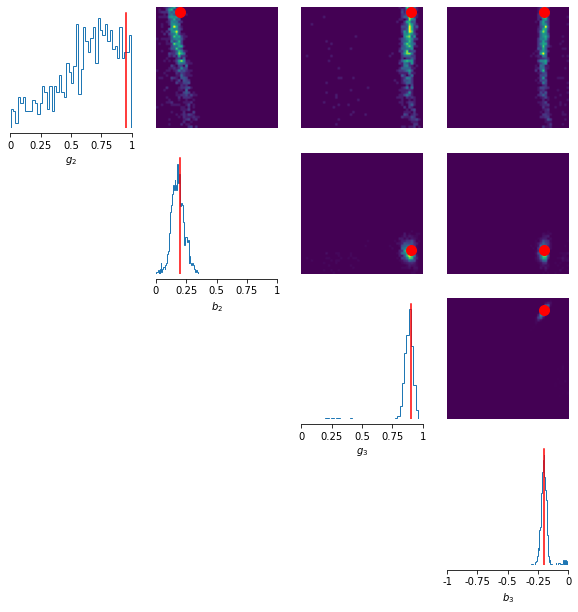} 
    \caption{Sequential neural posterior estimation with naive summary statistics, $10^4$ simulations 
    } 
    \label{fig:BH1NPEH} 
  \end{subfigure}
  \hfill 
  \begin{subfigure}[b]{0.48\linewidth}
    \centering
    \includegraphics[width=0.85\linewidth]{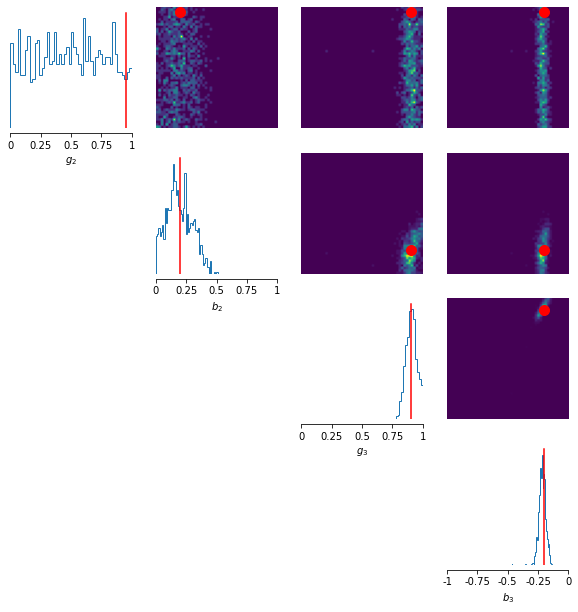} 
    \caption{Sequential neural posterior estimation with learned summary statistics, $10^4$ simulations 
    } 
    \label{fig:BH1NPEL} 
  \end{subfigure} 
  
  \vskip 0.15in
  \begin{subfigure}[b]{0.48\linewidth}
    \centering
    \includegraphics[width=0.85\linewidth]{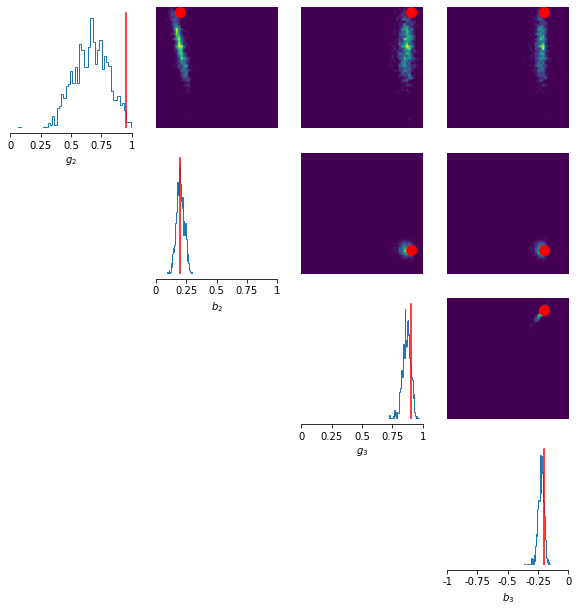} 
    \caption{Sequential neural ratio estimation with naive summary statistics, $10^4$ simulations 
    } 
    \label{fig:BH1NREH} 
  \end{subfigure}
  \hfill 
  \begin{subfigure}[b]{0.48\linewidth}
    \centering
    \includegraphics[width=0.85\linewidth]{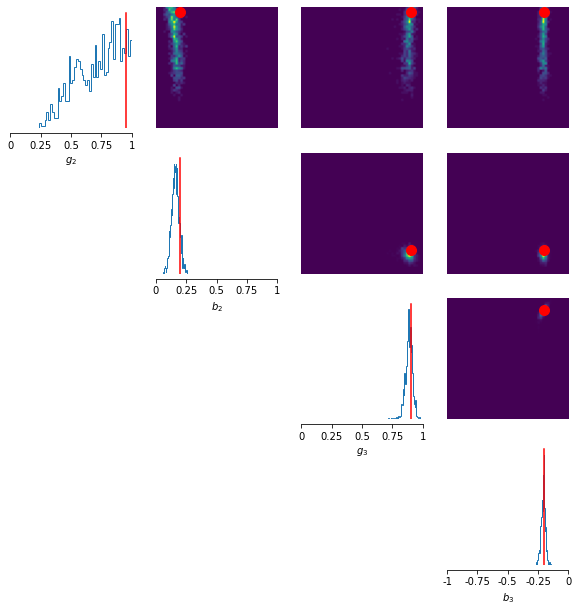} 
    \caption{Sequential neural ratio estimation with learned summary statistics, $10^4$ simulations 
    } 
    \label{fig:BH1NREL} 
  \end{subfigure} 
  
  \caption{(\textbf{Brock \& Hommes, parameter set 1}) Posteriors obtained with each posterior estimation method. The marginal posterior distributions are located on the diagonals, while the bivariate joint distributions for each parameter pair are located on the upper diagonal. Red lines/dots indicate the mean of the true posterior.}
  \label{fig:BH1} 
\end{figure}

In Figure \ref{fig:BH1}, we show the posteriors obtained using different posterior estimation methods: Figure \ref{fig:BH1True} shows the approximate ground-truth posterior density obtained with the true likelihood function and \gls{mh}; Figure \ref{fig:BH1G} shows the posterior obtained with \km{} and \gls{mh}; Figures \ref{fig:BH1NPEH} and \ref{fig:BH1NPEL} show the posterior estimated via \gls{snpe} with naive and learned summary statistics, respectively; and Figures \ref{fig:BH1NREH} and \ref{fig:BH1NREL} show the posterior obtained via \gls{snre} and naive and learned summary statistics, respectively, which were also sampled using \gls{mh}. Here, we use an embedding network consisting of two stacked Elman recurrent units with hidden state of size 32, followed by a single linear layer of size 16.  
In each figure, the marginals and joint bivariate densities are located on the diagonal and upper diagonal, respectively, while the red lines/dots locate the mean of the true posterior. 

We see from the approximate ground-truth in Figure \ref{fig:BH1True} that the marginal posteriors are sharply peaked on the generating parameters for $b_2, g_3$, and $b_3$, while the ground truth marginal for $g_2$ is shifted towards higher values. While these features are recovered reasonably well for Figures \ref{fig:BH1NPEH}--\ref{fig:BH1NREL} (the most notable exception being the posteriors for $g_2$), we see that they are recovered poorly with \km. This performance gap is also manifested in the \wasss and \mmds metrics reported in Table \ref{tab:BH}, where lower values indicate better estimates of the posterior. In summary, \kms both requires a far larger simulation budget to estimate a single posterior, while simultaneously generating worse estimates of that posterior. In contrast, \gls{snpe} and \gls{snre} is able to achieve superior estimates of the ground-truth posterior distribution, despite the 10-fold reduction in the simulation budget they have been afforded.

\begin{table}
\centering
\begin{tabularx}{\textwidth}{c c *{5}{Y}}
\toprule[0.3ex]
\textbf{Parameter set} & \textbf{Metric} & \multicolumn{5}{c}{\textbf{Estimation method}}\\
\cmidrule(l){3-7}
{} & {}  & \kms & \gls{snpe} & \gls{snpe}* & \gls{snre} & \gls{snre}* \\
\midrule \\
\multirow{2}{*}{\textbf{1}} & \textsc{Wasserstein} & 0.690 & 0.477 & 0.336 & \textbf{0.241} & \emph{0.299} \\
 & \textsc{mmd}         & 1.015 & 0.789 & \emph{0.552} & \textbf{0.451} & 0.781 \\
 
 \midrule
 
\multirow{2}{*}{\textbf{2}} & \textsc{Wasserstein} & 0.304 & \textbf{0.154} & 0.306 & \emph{0.164} & 0.291 \\
 & \textsc{mmd}        & 0.127  & \textbf{0.036} & 0.133 & \emph{0.041} & 0.118 \\
 
\bottomrule\\
\textbf{Simulation budget} & {} & $1.5\times 10^5$ & $10^4$ & $10^4$ & $10^4$ & $10^4$\\
\bottomrule[0.3ex]
\end{tabularx}
\begin{minipage}{\textwidth}
\caption{\textbf{(Brock \& Hommes)} Discrepancies between the approximate ground-truth posterior and the posteriors estimated with \km, \gls{snpe}, and \gls{snre}. \textbf{Bold} and \emph{italics} indicate best and second-best, respectively. For the neural methods, * indicates that the naive hand-crafted summary statistics described in the main text were used, otherwise summary statistics are learned from the simulated data.}\label{tab:BH}
\end{minipage}
\end{table}

\subsubsection{Parameter set 2}\label{sec:bh_noisy}

We now take $\beta = 10$ and consider the task of estimating the posterior density $p(\bth \mid \by)$, where $\by := (y_1, y_2, \dots, y_T) \sim p(\bx \mid \bth^{*})$, $T=100$, and $\bth^{*} := (g_2^{*}, b_2^{*}, g_3^{*}, b_3^{*}) = (-0.7, -0.4, 0.5, 0.3)$. In this case, we use the following priors: $g_2, b_2 \sim \mathcal{U}(-1,0)$ and $g_3, b_3 \sim \mathcal{U}(0,1)$.

\begin{figure} 
  \begin{subfigure}[b]{0.48\linewidth}
    \centering
    \includegraphics[width=0.85\linewidth]{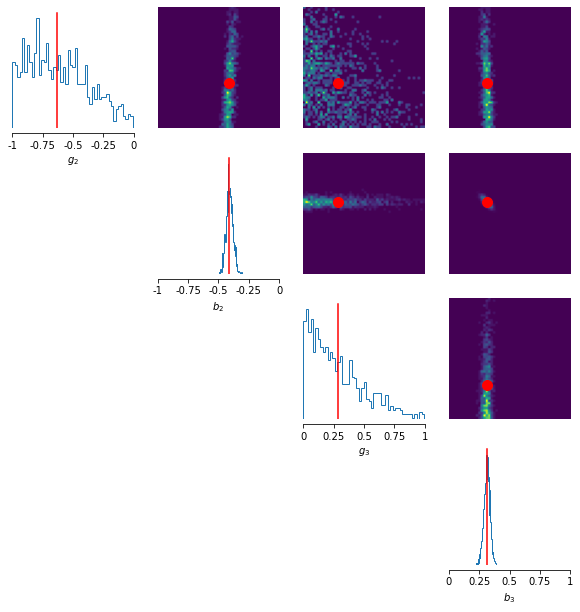} 
    \caption{True posterior} 
    \label{fig:BH2True} 
    \vspace{4ex}
  \end{subfigure}
  \hfill
  \begin{subfigure}[b]{0.48\linewidth}
    \centering
    \includegraphics[width=0.85\linewidth]{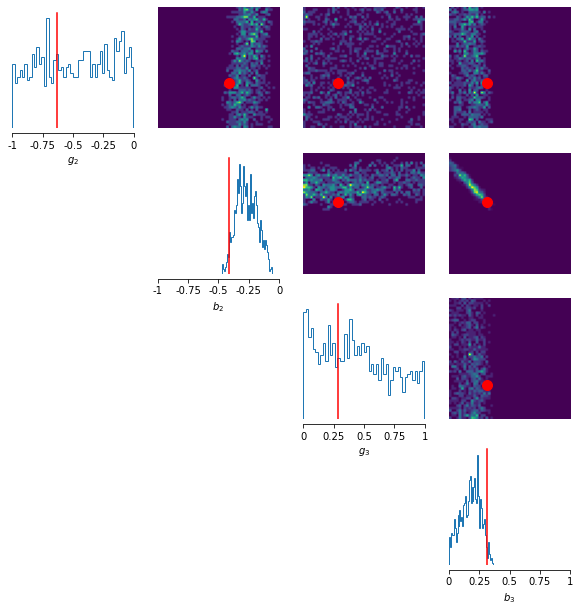} 
    \caption{
             \textsc{kde} posterior, $10^5$ simulations} 
    \label{fig:BH2G} 
    \vspace{4ex}
  \end{subfigure} 
  
  \vskip -0.1in
  \begin{subfigure}[b]{0.48\linewidth}
    \centering
    \includegraphics[width=0.85\linewidth]{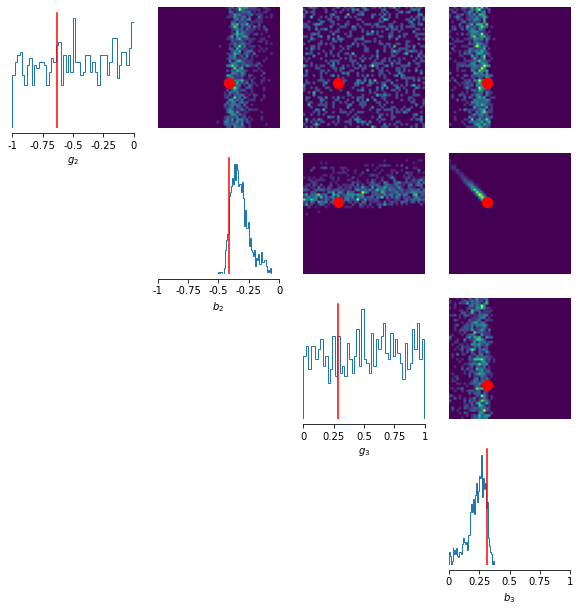} 
    \caption{Sequential neural posterior estimation with naive summary statistics, $10^4$ simulations} 
    \label{fig:BH2NPEH} 
  \end{subfigure}
  \hfill 
  \begin{subfigure}[b]{0.48\linewidth}
    \centering
    \includegraphics[width=0.85\linewidth]{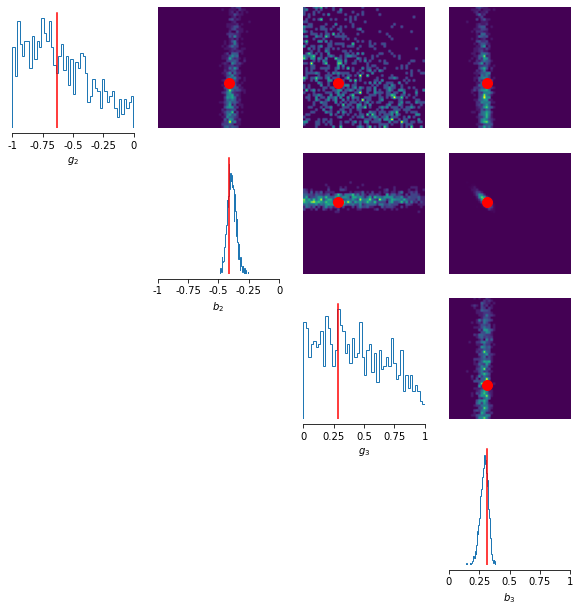} 
    \caption{Sequential neural posterior estimation with learned summary statistics, $10^4$ simulations} 
    \label{fig:BH2NPEL} 
  \end{subfigure}
  
  \vskip 0.15in
  \begin{subfigure}[b]{0.48\linewidth}
    \centering
    \includegraphics[width=0.85\linewidth]{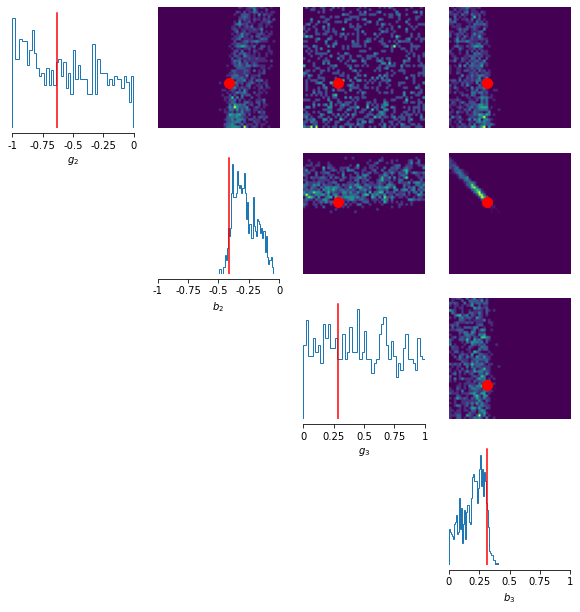} 
    \caption{Sequential neural ratio estimation with naive summary statistics, $10^4$ simulations
    } 
    \label{fig:BH2NREH} 
  \end{subfigure}
  \hfill 
  \begin{subfigure}[b]{0.48\linewidth}
    \centering
    \includegraphics[width=0.85\linewidth]{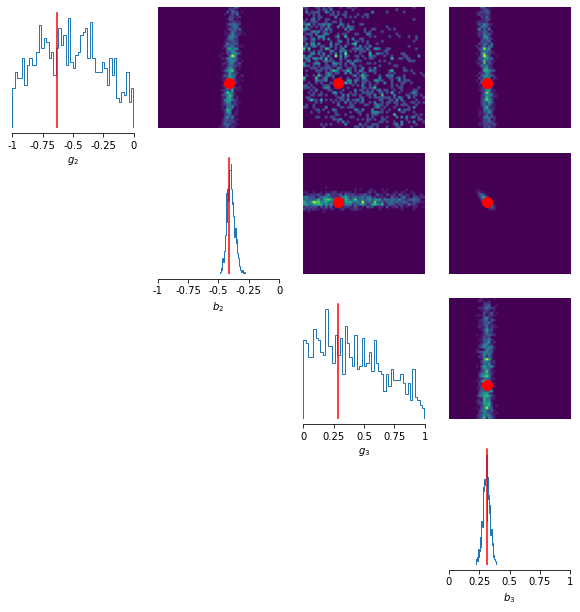} 
    \caption{Sequential neural ratio estimation with learned summary statistics, $10^4$ simulations
    } 
    \label{fig:BH2NREL} 
  \end{subfigure} 
  \caption{(\textbf{Brock \& Hommes, parameter set 2}) Posteriors obtained with each posterior estimation method. The marginal posterior distributions are located on the diagonals, while the bivariate joint distributions for each parameter pair are located on the upper diagonal. Red lines/dots indicate the mean of the true posterior.}
  \label{fig:BH2} 
\end{figure}

In Figure \ref{fig:BH2}, we show the posteriors obtained using different posterior estimation methods: Figure \ref{fig:BH2True} shows the approximate ground-truth posterior density obtained with the true likelihood function and \gls{mh}; Figure \ref{fig:BH2G} shows the posterior obtained with \km{} and \gls{mh}; Figures \ref{fig:BH2NPEH} and \ref{fig:BH2NPEL} show the posterior estimated via \gls{snpe} with naive and learned summary statistics, respectively; and Figures \ref{fig:BH2NREH} and \ref{fig:BH2NREL} show the posterior obtained via \gls{snre} and naive and learned summary statistics, respectively, which were also sampled with \gls{mh}. In this experiment, we now use an embedding network consisting of two stacked \glspl{gru} with hidden state of size 32, followed by a single linear layer of size 16. This results in an embedding network with approximately 10,000 trainable parameters.

We see from the approximate ground-truth in Figure \ref{fig:BH2True} that the marginal posteriors for $b_2$ and $b_3$ remain sharply peaked on the generating parameters, but that the ground truth marginals for $g_2$ and $g_3$ are diffuse and sloped. The diffuseness of the posterior with respect to these two dimensions is also visually apparent in the joint density plots in the upper diagonal. Upon inspection of the estimated posteriors, we see that the overall shape is recovered well by \gls{snpe} and \gls{snre} with learned summary statistics, but less accurately by \kms and \gls{snpe} and \gls{snre} when using the naive hand-crafted summary statistics described in Section \ref{sec:BH}. This is once again reflected by the \wasss and \mmds scores, reported in Table \ref{tab:BH}, in which we see significantly decreased values for \gls{snpe} and \gls{snre} with learned summary statistics with respect to the alternatives, despite a 10-fold decrease in the simulation budget they have been afforded. It is nonetheless noteworthy that even with the naive hand-crafted summary statistics, \gls{snpe} and \gls{snre} achieve comparable and slightly favourable performance than \km, again with a 10-fold decrease in the allotted simulation budget.

\subsection{Example 2: Multivariate Geometric Brownian Motion}

In Section \ref{sec:BH}, we demonstrated that \gls{npe} and \gls{nre} can be used to generate Bayesian parameter posteriors that better match the ground truth posterior than \kms for the univariate Brock \& Hommes model, despite the fact that the latter method entailed 10 times as many simulations as the former two methods based on neural networks. In this section, we seek to demonstrate that this remains the case even for models that generate multivariate time-series as output.

To this end, we consider a \gls{mvgbm}, which is used in a variety of applications in financial time-series modelling. The model is a  stochastic differential equation in which each of the $d$ components, labelled $i \in \lbrace{1, 2, \dots, d\rbrace}$, of the path $\bX_t = \left(X^1_t, X^2_t, \dots, X^d_t\right) \in \mathbb{R}^d_{+}$ evolve according to
\begin{equation}
    \textrm{ d}X^i_t = X^i_t \left(\left[b_i - \frac{1}{2}\sum_{j=1}^{d} \sigma_{ij}^2 \right] \textrm{ d}t + \sum_{j=1}^{d} \sigma_{ij} \textrm{ d}W^{j}_t \right),
\end{equation}
where the $b_i$ are drift coefficients, the $\sigma_{ij}$ are volatility coefficients, and $W^i_t$ is a Brownian motion.

We consider the case of $d=3$ and the task of estimating the posterior for the parameters $\bth = (b_1, b_2, b_3)$ given an observation $\by \sim p\left(\bx \mid \bth^{*}\right)$ of $T=100$ points spaced equally with spacing $\Delta t = 1/(T-1)$, where $\bth^{*} = \left(0.2, -0.5, 0.0\right)$. We take priors $b_i \sim \mathcal{U}(-1, 1)$ for each $i = 1, 2, 3$. We note that this model once again permits both exact simulations and samples from the exact posterior since the transition density is once again tractable\footnote{Assuming that $\sigma$ is of full rank.} and can be written as
\begin{equation}
    \bX_{t + \Delta t} \sim \mathcal{N}\left(\bX_t + \left(\bth - \boldsymbol{\gamma} \right)\Delta t,\ \sigma \sigma^T \Delta t\right),
\end{equation}
where 
\begin{equation}
    \boldsymbol{\gamma} = \frac{1}{2}\left[ \sum_{j=1}^{d} \sigma_{1j}^2,\ \ \sum_{j=1}^{d} \sigma_{2j}^2,\ \ \sum_{j=1}^{d} \sigma_{3j}^2\right]'.
\end{equation}
In our experiments, we take
\begin{equation}
    \sigma = \begin{pmatrix}
    0.5 & 0.1 & 0.0\\
    0.0 & 0.1 & 0.3\\
    0.0 & 0.0 & 0.2
    \end{pmatrix}.
\end{equation}

\begin{figure} 
  \begin{subfigure}[b]{0.475\linewidth}
    \centering
    \includegraphics[width=\linewidth]{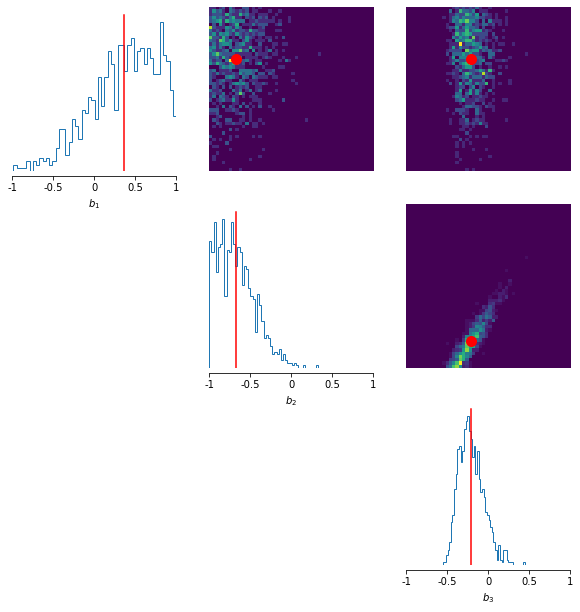}
    \caption{True posterior} 
    \label{fig:MVGBM_True} 
    \vspace{4ex}
  \end{subfigure}
  \hfill
  \begin{subfigure}[b]{0.475\linewidth}
    \centering
    \includegraphics[width=\linewidth]{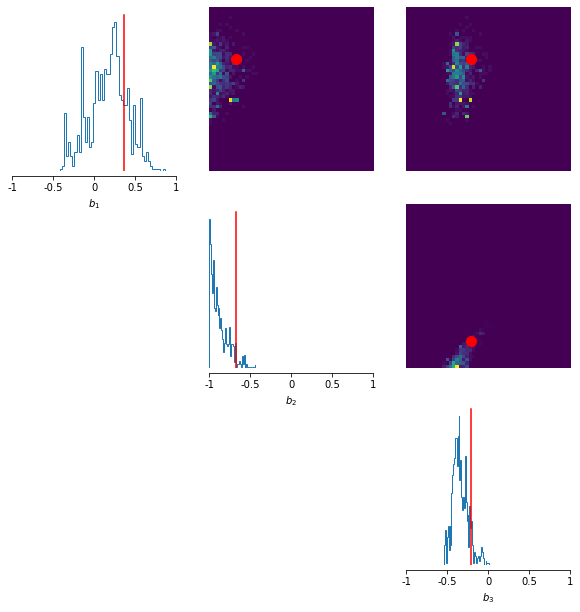}
    \caption{
             \km{} posterior, $10^6$ simulations} 
    \label{fig:mvgbm_kde} 
    \vspace{4ex}
  \end{subfigure} 
  
  \begin{subfigure}[b]{0.475\linewidth}
    \centering
    \includegraphics[width=\linewidth]{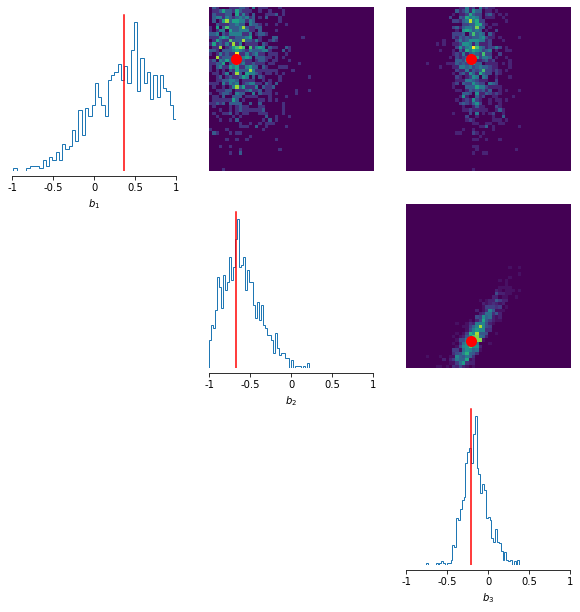}
    \caption{Neural posterior estimation with automatically learned summary statistics, $10^3$ simulations} 
    \label{fig:mvgbm_maf} 
  \end{subfigure}
  \hfill 
  \begin{subfigure}[b]{0.475\linewidth}
    \centering
    \includegraphics[width=\linewidth]{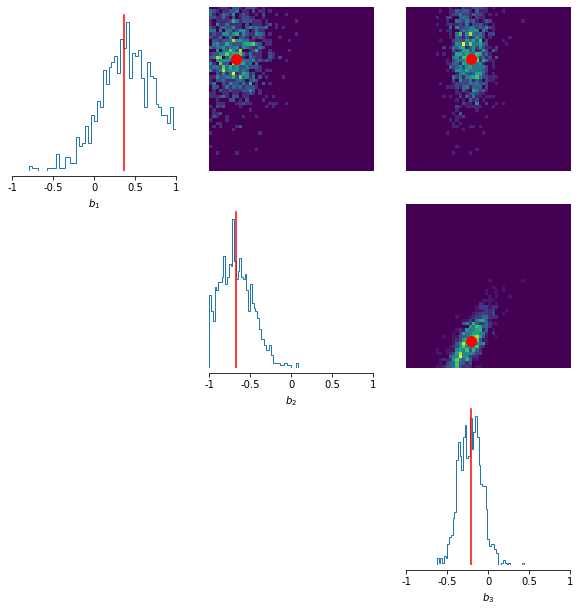} 
    \caption{Neural ratio estimation with automatically learned summary statistics, $10^3$ simulations} 
    \label{fig:mvgbm_resnet} 
  \end{subfigure}
  \caption{(\textbf{Multivariate geometric Brownian motion}) Posteriors obtained with each estimation method. The marginal posterior distributions are located on the diagonals, while the bivariate joint distributions for each parameter pair are located on the upper diagonal. Red lines/dots indicate the mean of the true posterior.}
\end{figure}

In Figure \ref{fig:MVGBM_True} we show the approximate ground truth posterior obtained with \gls{mh} and the transition density described above, while in Figures \ref{fig:mvgbm_kde}, \ref{fig:mvgbm_maf}, and \ref{fig:mvgbm_resnet} we show the posteriors obtained with \km{} and \gls{mh}, \gls{npe} with learned summary statistics, and \gls{nre} with \gls{mh} also with learned summary statistics, respectively. The corresponding simulation budgets are $10^6$, $10^3$, and $10^3$, respectively. To learn the summary statistics, we once again use the embedding network described in Section \ref{sec:bh_noisy} and learn summary statistics and the density (ratio) estimator concurrently.

We see that the approximate ground-truth posteriors are relatively diffuse, with peaks approximately coinciding with the true posterior mean, shown with red lines/dots. The shape and degree of diffuseness is captured accurately by \gls{npe} and \gls{nre}. In contrast, the posterior obtained with \kms is insufficiently diffuse and biased, and thus a significantly worse estimate of the ground-truth posterior. These observations are corroborated by the corresponding \wasss and \mmds metrics, reported in Table \ref{tab:mvgbm} along with the corresponding simulation budgets. In summary, \gls{npe} and \gls{nre} achieve significantly more accurate posterior estimates here with a 1000-fold decrease in the simulation budget, and are able to flexibly accommodate multivariate time-series as input for the inference problem.

\begin{table}
\centering
\begin{tabularx}{0.6\textwidth}{c *{3}{Y}}
\toprule[0.3ex]
\textbf{Metric} & \multicolumn{3}{c}{\textbf{Estimation method}}\\
\cmidrule(l){2-4}
 {} & \kms & \gls{npe} & \gls{nre} \\
\midrule 
 & & & \\
 \textsc{Wasserstein} & 0.364 & \textbf{0.099} & \emph{0.107} \\
 \textsc{mmd}         & 0.137 & \emph{0.005} & \textbf{0.004} \\
\bottomrule\\
\textbf{Simulation budget} & $10^6$ & $10^3$ & $10^3$\\
\bottomrule[0.3ex]
\end{tabularx}
\begin{minipage}{0.9\textwidth}
\caption{\textbf{(Multivariate geometric Brownian motion)} Discrepancies between the approximate ground-truth posterior and the posteriors estimated with \gls{npe}, \gls{nre}, and \km. Smaller values indicate more accurate posteriors; \textbf{bold} and \emph{italics} indicate best and second-best, respectively.}\label{tab:mvgbm}
\end{minipage}
\end{table}

\section{Validating approximate Bayesian inference}

As previously mentioned, a major benefit of the Bayesian inferential paradigm is the fact that uncertainty quantification is a built-in feature captured via the posterior distribution. Its utility relies, however, on the ability to capture the correct degree of diffuseness in the posterior, such that the uncertainty quantification in the recovered posterior is meaningful. This presents a challenge in approximate Bayesian inference, since by definition the experimenter does not know the correct shape of the posterior when exact inference is intractable. In this section, we address this issue by outlining one widely used approach to verifying the accuracy of approximate Bayesian inference pipelines in the absence of any ground-truth posterior densities. 

\subsection{Simulation-based calibration}\label{sec:sbc}

Simulation-based calibration is a general purpose method for validating approximate Bayesian inference pipelines. The core idea is to use the fact that the \emph{data-averaged posterior} should be identical to the prior distribution $p(\bth)$. That is, a perfect Bayesian inference pipeline should uphold the following equality:
\begin{equation}
    \int_{\mathcal{Y}\times \bTh} p(\bth \mid \by) p(\by \mid \tilde{\bth}) p(\tilde{\bth})\, {\textrm{d} \tilde{\bth}}\, {\textrm{d} \by} = p(\bth).
\end{equation}
A deviation from this equality signifies some error in the posterior sampling procedure; thus, it has been proposed by \citet{talts2020validating} that testing how close our Bayesian workflow is to satisfying this equality is a test of the accuracy of the Bayesian pipeline. This may be performed by repeating the following steps $P$ times:
\begin{enumerate}
    \item Generate $\tilde{\bth} \sim p(\bth)$ from the prior distribution
    \item Generate $\tilde{\by} \sim p(\by \mid \tilde{\bth})$ from the likelihood function (i.e. the simulator)
    \item Generate $L$ uncorrelated posterior samples $\left\{\bth_{i}\right\}_{i=1}^{L} \sim p(\bth \mid \tilde{\by})$
    \item Compute and store the rank statistic $r(\left\{\bth_{i}\right\}_{i=1}^{L}, \tilde{\bth}) \in \left\{0, \dots, L\right\}$ for $\tilde{\bth}$ within $\left\{\bth_{i}\right\}_{i=1}^{L} \cup \left\{\tilde{\bth}\right\}$
\end{enumerate}
It can be shown \citep[Theorem 1,][]{talts2020validating} that the rank statistics obtained as above should follow a discrete Uniform distribution on $\left\{0, \dots, L\right\}$ for any joint distribution $p(\by \mid \bth)p(\bth)$ for one-dimensional $\bth$, while for $d$-dimensional $\bth$ each of the $d$ components' rank histograms should be uniform. Inspecting the discrepancy between the true distribution of rank statistics and the desired Uniform distribution thus gives an indication of the accuracy of the Bayesian pipeline; furthermore and conversely, particular patterns of deviation from the desired uniformity are interpretable and provide insight into the specific way in which the obtained posteriors are inaccurate \citep{talts2020validating}. Performing this procedure, and inspecting the resultant rank histograms, is referred to as performing \gls{sbc}.

While this procedure is in principle possible for \km, the computational burden is immense even for cheap simulation models, and becomes infeasible for large-scale \gls{abm}s. This is because the number of simulations required to generate $n$ samples from the posterior would, in general, be at least\footnote{In practice, this number may be larger due to the necessity of performing trial runs to estimate the parameters of the proposal distribution.} $n R$, such that the total number of simulations required to perform \gls{sbc} for \kms increases to $P  n  R$. Even for a moderately sized \gls{sbc} task with $P \simeq 10^3$ and the most optimistic $R=1$, the total number of simulations required can easily reach the order of $10^8$, since it is typically necessary to take $n$ to be many hundreds of thousands to obtain a reasonably accurate estimate of the posterior. In fact, larger values of $n$ are likely to be necessary, since \gls{abm}s are usually highly parameterised and thus can have large parameter spaces.

In contrast, such an approach to verifying the accuracy of posteriors derived from \gls{npe} and \gls{nre} is feasible due to the fact that such approaches involve the prior training of a global density (ratio) estimator, obviating any simulations during the inference (posterior sampling) phase. Thus the strength of \gls{npe} and \gls{nre} derives from not only their enhanced performance and improved simulation-efficiency, but also from the fact that such methods permit accuracy checks such as \gls{sbc} that are otherwise infeasible for alternative parameter estimation techniques when the simulator is expensive, as is often the case for large-scale \glspl{abm}.

\subsubsection{Example: \citet{franke2012structural}}

To further demonstrate the ability of \gls{npe} and \gls{nre} to recover accurate posterior estimates and informative summary statistics in a highly automated manner, we perform \gls{sbc} using a posterior and density ratio estimator trained on the Franke \& Westerhoff model \citep{franke2012structural}, which has previously been used in benchmarking experiments for \gls{abm} estimation methods \citep{platt2021bayesian}. In particular, we consider the model version referred to as the ``Wealth \& Predisposition'' model which, similarly to the Brock \& Hommes model (see Section \ref{sec:BH}), may be written as a system of coupled equations which models the asset price dynamics that result from a heterogeneous system of traders:
\begin{align}\label{eq:fw1}
    p_t &= p_{t-1} + \mu\left(n_{t-1}^f d_{t-1}^f + n_{t-1}^c d_{t-1}^c\right),\\
    d_t^f &= \phi\left(p^{*} - p_{t}\right) + \sigma_f \epsilon_{t}^f,\\
    d_t^c &= \chi\left(p_{t} - p_{t-1}\right) + \sigma_c \epsilon_{t}^c,\\
    n_t^f &= 1 - n_t^c = \frac{1}{1 + \exp{\left(-\beta a_{t-1}\right)}},\\
    a_t &= \alpha_w \left(w^f_t - w^c_t\right) + \alpha_0,\\
     w^j_t &= \eta w^j_{t-1} + \left(1 - \eta \right)g^j_t,\ \ j\in \lbrace{f, c\rbrace}\\   \label{eq:fwn}
    g^j_t &= \left[e^{p_t} - e^{p_{t-1}}\right]d^j_{t-2},\ \ j\in \lbrace{f, c\rbrace}
\end{align}
where the $\epsilon^j_t \sim \mathcal{N}(0,1)$ are standard normal random variables and $j = \lbrace{f, c\rbrace}$ labels fundamentalist and chartist traders, respectively. We take as output from the model the time series of log returns $r_t = p_t - p_{t-1}$, and assume the task of training a global, amortised density and density ratio estimator via \gls{npe} and \gls{nre}, respectively, using $10^4$ simulations of length $T=100$ from the simulator defined by Equations \eqref{eq:fw1}--\eqref{eq:fwn} with the following parameter settings: $\mu = 0.01$, $\beta = 1$, $\phi = 1$, $\chi = 0.9$, $\alpha_0 = 2.1$, $\sigma_f = 0.752$. The parameter vector for which we seek the posterior and density ratio estimators is taken to be $\bth = \left(\alpha_w, \eta, \sigma_c\right)$, and we obtain \emph{iid} samples from the posterior associated with the density ratio estimator via a sampling-importance-resampling scheme described in Appendix \ref{app:sir}. (Samples may be drawn \emph{iid} from the \gls{npe} posterior by construction.)

\begin{figure} 
\centering
  \begin{subfigure}[b]{0.75\linewidth}
    \centering
    \includegraphics[width=\linewidth]{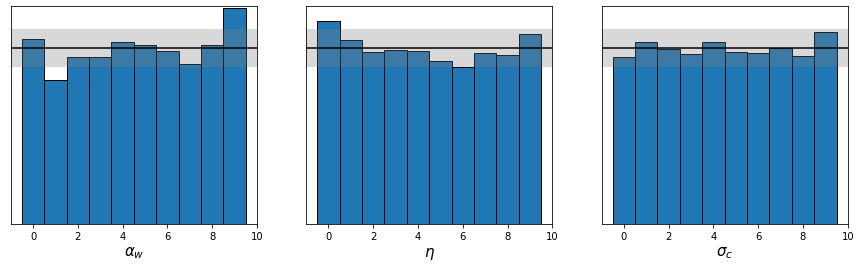}
    \caption{Simulation-based calibration histogram for neural posterior estimation.} 
    \label{fig:fw_wp_maf} 
    \vspace{4ex}
  \end{subfigure}
  
  \begin{subfigure}[b]{0.75\linewidth}
    \centering
    \includegraphics[width=\linewidth]{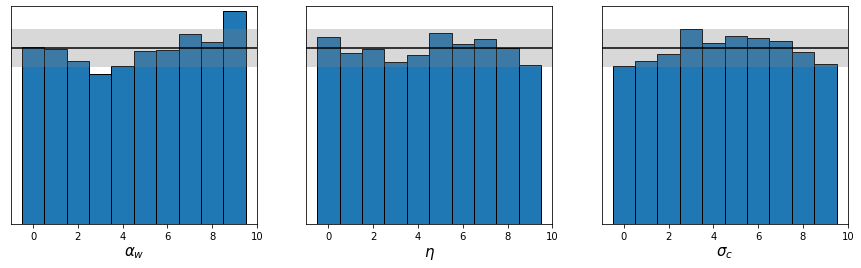}
    \caption{Simulation-based calibration histogram for neural density ratio estimation.} 
    \label{fig:fw_wp_nre} 
  \end{subfigure}

  \caption{(\textbf{Franke \& Westerhoff}) Rank histograms generated according to the simulation-based calibration procedure in Section \ref{sec:sbc} using the trained posterior density estimator (top) and density ratio estimator (bottom).}\label{fig:sbc_fw}
\end{figure}

We show in Figure \ref{fig:sbc_fw} the rank histograms obtained via the simulation-based calibration procedure in Section \ref{sec:sbc} with \gls{npe} (Figure \ref{fig:fw_wp_maf}) and \gls{nre} (Figure \ref{fig:fw_wp_nre}). To construct these histograms, we take $P = 5,000$ and the following uniform priors: $\alpha_w \sim \mathcal{U}\left(0, 15000\right)$, $\eta \sim \mathcal{U}(0, 1)$, and $\sigma_c \sim \mathcal{U}\left(0, 5\right)$. The black line and gray band denotes the expected height and the expected variation in the heights, respectively, of the bars at this $P$ and with this number of bins. We see that the bars tend to lie within the expected range, although some notable deviations exist: for example, the rank histograms for $\alpha_w$ for both \gls{npe} and \gls{nre} exhibit a minor bias towards larger rank values, which suggests that the marginal posteriors for $\alpha_w$ in both cases is slightly biased towards lower values on average than the true posteriors \citep{talts2020validating}. However, we emphasise that a major benefit of \gls{npe} and \gls{nre} is that they allow the modeller to make statements of this sort in the first place, whereas the number of simulations required to make such statements in the case of more traditional techniques such as \kms would quickly become prohibitively large.

\section{Conclusion}

\glsresetall

In this paper, we investigated the use of two recently developed approaches to Bayesian parameter estimation for intractable simulation models: 
\gls{npe} for approximating the posterior density directly, and \gls{nre} for approximating the likelihood-to-evidence ratio. We motivated the use of these parameter inference methods as simulation-efficient black-box discriminative approaches to Bayesian estimation for complex time-series simulators such as agent-based models in economics, which contrast with existing methods that (a) entail an often prohibitively large simulation burden, (b) can make restrictive assumptions about the form of the simulator, for example by ignoring important structural features such as temporal dependencies, and (c) are inherently generative, and thus assume a task that is typically more difficult than discrimination alone. We argue that this latter point is likely to be particularly pertinent for the case of agent-based models in the social sciences, since they are known to be able to generate complex and stochastic non-equilibrium dynamics that can be difficult to model and generate.

We further reviewed existing alternatives, and benchmarked \gls{npe} and \gls{nre} against the most popular of these. In these examples, we see that \gls{npe} and \gls{nre} generally achieve superior recovery of the approximate ground-truth posterior distributions, despite requiring simulation budgets that are orders of magnitude lower than traditional alternatives. In addition, we demonstrated that a verification of the accuracy of the density (ratio) estimators is possible with the introduced methods via \gls{sbc}, due to the amortisation of the estimators. In contrast, such checks would necessitate an unreasonably large number of simulations for existing methods in the literature for Bayesian estimation of economic simulation models, for example the methods discussed by \citet{grazzini2017bayesian}. For these reasons, we argue that such simulation-efficient black-box Bayesian inference as \gls{npe} and \gls{nre} may enable economists and social scientists alike to more readily exploit the potential benefits of the agent-based modelling paradigm. 

\clearpage

\bibliography{references}

\begin{thebibliography}{79}
\providecommand{\natexlab}[1]{#1}
\providecommand{\url}[1]{\texttt{#1}}
\expandafter\ifx\csname urlstyle\endcsname\relax
  \providecommand{\doi}[1]{doi: #1}\else
  \providecommand{\doi}{doi: \begingroup \urlstyle{rm}\Url}\fi

\bibitem[Abadi et~al.(2016)Abadi, Barham, Chen, Chen, Davis, Dean, Devin,
  Ghemawat, Irving, Isard, et~al.]{abadi2016tensorflow}
Mart{\'\i}n Abadi, Paul Barham, Jianmin Chen, Zhifeng Chen, Andy Davis, Jeffrey
  Dean, Matthieu Devin, Sanjay Ghemawat, Geoffrey Irving, Michael Isard, et~al.
\newblock Tensorflow: A system for large-scale machine learning.
\newblock In \emph{12th $\{$USENIX$\}$ symposium on operating systems design
  and implementation ($\{$OSDI$\}$ 16)}, pages 265--283, 2016.

\bibitem[Alsing et~al.(2019)Alsing, Charnock, Feeney, and
  Wandelt]{alsing2019fast}
Justin Alsing, Tom Charnock, Stephen Feeney, and Benjamin Wandelt.
\newblock {Fast likelihood-free cosmology with neural density estimators and
  active learning}.
\newblock \emph{Monthly Notices of the Royal Astronomical Society},
  488\penalty0 (3):\penalty0 4440--4458, 07 2019.
\newblock ISSN 0035-8711.
\newblock \doi{10.1093/mnras/stz1960}.
\newblock URL \url{https://doi.org/10.1093/mnras/stz1960}.

\bibitem[Andrieu and Roberts(2009)]{andrieu2009pseudo}
Christophe Andrieu and Gareth~O Roberts.
\newblock The pseudo-marginal approach for efficient monte carlo computations.
\newblock \emph{The Annals of Statistics}, 37\penalty0 (2):\penalty0 697--725,
  2009.

\bibitem[Arjovsky et~al.(2017)Arjovsky, Chintala, and Bottou]{wgan}
Martin Arjovsky, Soumith Chintala, and L{\'e}on Bottou.
\newblock {W}asserstein generative adversarial networks.
\newblock In Doina Precup and Yee~Whye Teh, editors, \emph{Proceedings of the
  34th International Conference on Machine Learning}, volume~70 of
  \emph{Proceedings of Machine Learning Research}, pages 214--223. PMLR, 06--11
  Aug 2017.
\newblock URL \url{https://proceedings.mlr.press/v70/arjovsky17a.html}.

\bibitem[Beaumont(2010)]{beaumont2010approximate}
Mark~A Beaumont.
\newblock Approximate bayesian computation in evolution and ecology.
\newblock \emph{Annual review of ecology, evolution, and systematics},
  41:\penalty0 379--406, 2010.

\bibitem[Beaumont et~al.(2002)Beaumont, Zhang, and
  Balding]{beaumont2002approximate}
Mark~A Beaumont, Wenyang Zhang, and David~J Balding.
\newblock Approximate bayesian computation in population genetics.
\newblock \emph{Genetics}, 162\penalty0 (4):\penalty0 2025--2035, 2002.

\bibitem[Beaumont et~al.(2009)Beaumont, Cornuet, Marin, and
  Robert]{beaumont2009adaptive}
Mark~A Beaumont, Jean-Marie Cornuet, Jean-Michel Marin, and Christian~P Robert.
\newblock Adaptive approximate bayesian computation.
\newblock \emph{Biometrika}, 96\penalty0 (4):\penalty0 983--990, 2009.

\bibitem[Bernton et~al.(2019)Bernton, Jacob, Gerber, and Robert]{Bernton2019}
Espen Bernton, Pierre~E. Jacob, Mathieu Gerber, and Christian~P. Robert.
\newblock {Approximate Bayesian computation with the Wasserstein distance}.
\newblock \emph{Journal of the Royal Statistical Society. Series B: Statistical
  Methodology}, 81\penalty0 (2):\penalty0 235--269, 2019.
\newblock ISSN 14679868.
\newblock \doi{10.1111/rssb.12312}.

\bibitem[Bishop(1994)]{bishop1994mixture}
Christopher~M Bishop.
\newblock Mixture density networks.
\newblock 1994.

\bibitem[Blum et~al.(2013)Blum, Nunes, Prangle, and
  Sisson]{blum2013comparative}
Michael~GB Blum, Maria~Antonieta Nunes, Dennis Prangle, and Scott~A Sisson.
\newblock A comparative review of dimension reduction methods in approximate
  bayesian computation.
\newblock \emph{Statistical Science}, 28\penalty0 (2):\penalty0 189--208, 2013.

\bibitem[Bornn et~al.(2017)Bornn, Pillai, Smith, and Woodard]{bornn2017use}
Luke Bornn, Natesh~S Pillai, Aaron Smith, and Dawn Woodard.
\newblock The use of a single pseudo-sample in approximate bayesian
  computation.
\newblock \emph{Statistics and Computing}, 27\penalty0 (3):\penalty0 583--590,
  2017.

\bibitem[Brehmer et~al.(2018)Brehmer, Cranmer, Louppe, and
  Pavez]{brhemer2018constraining}
Johann Brehmer, Kyle Cranmer, Gilles Louppe, and Juan Pavez.
\newblock Constraining effective field theories with machine learning.
\newblock \emph{Phys. Rev. Lett.}, 121:\penalty0 111801, Sep 2018.
\newblock \doi{10.1103/PhysRevLett.121.111801}.
\newblock URL \url{https://link.aps.org/doi/10.1103/PhysRevLett.121.111801}.

\bibitem[Briol et~al.(2019)Briol, Barp, Duncan, and Girolami]{Briol2019}
Fran{\c{c}}ois~Xavier Briol, Alessandro Barp, Andrew~B. Duncan, and Mark
  Girolami.
\newblock {Statistical inference for generative models with maximum mean
  discrepancy}.
\newblock \emph{arXiv}, pages 1--57, 2019.

\bibitem[Brock and Hommes(1998)]{BROCK19981235}
William~A. Brock and Cars~H. Hommes.
\newblock Heterogeneous beliefs and routes to chaos in a simple asset pricing
  model.
\newblock \emph{Journal of Economic Dynamics and Control}, 22\penalty0
  (8):\penalty0 1235--1274, 1998.
\newblock ISSN 0165-1889.
\newblock \doi{https://doi.org/10.1016/S0165-1889(98)00011-6}.
\newblock URL
  \url{https://www.sciencedirect.com/science/article/pii/S0165188998000116}.

\bibitem[Chen et~al.(2020)Chen, Zhang, Gutmann, Courville, and Zhu]{Chen2020}
Yanzhi Chen, Dinghuai Zhang, Michael Gutmann, Aaron Courville, and Zhanxing
  Zhu.
\newblock {Neural Approximate Sufficient Statistics for Implicit Models}.
\newblock pages 1--14, 2020.
\newblock URL \url{http://arxiv.org/abs/2010.10079}.

\bibitem[Darmois(1935)]{darmois1935sufficient}
G~Darmois.
\newblock Sur les lois de probabilites a estimation exhaustive.
\newblock In \emph{C. R. Acad. Sci. Paris (in French)}, volume 200, page
  1265–1266, 1935.

\bibitem[Diggle and Gratton(1984)]{diggle1984monte}
Peter~J Diggle and Richard~J Gratton.
\newblock Monte carlo methods of inference for implicit statistical models.
\newblock \emph{Journal of the Royal Statistical Society: Series B
  (Methodological)}, 46\penalty0 (2):\penalty0 193--212, 1984.

\bibitem[Doucet et~al.(2015)Doucet, Pitt, Deligiannidis, and
  Kohn]{doucet2015efficient}
Arnaud Doucet, Michael~K Pitt, George Deligiannidis, and Robert Kohn.
\newblock Efficient implementation of markov chain monte carlo when using an
  unbiased likelihood estimator.
\newblock \emph{Biometrika}, 102\penalty0 (2):\penalty0 295--313, 2015.

\bibitem[Durkan et~al.(2020)Durkan, Murray, and Papamakarios]{Durkan2020}
Conor Durkan, Iain Murray, and George Papamakarios.
\newblock {On Contrastive Learning for Likelihood-free Inference}.
\newblock In Hal~Daumé III and Aarti Singh, editors, \emph{Proceedings of the
  37th International Conference on Machine Learning}, volume 119 of
  \emph{Proceedings of Machine Learning Research}, pages 2771--2781. PMLR,
  13--18 Jul 2020.
\newblock URL \url{http://proceedings.mlr.press/v119/durkan20a.html}.

\bibitem[Dyer et~al.(2021{\natexlab{a}})Dyer, Cannon, and
  Schmon]{dyer2021approximate}
Joel Dyer, Patrick Cannon, and Sebastian~M Schmon.
\newblock {Approximate Bayesian Computation with Path Signatures}.
\newblock \emph{arXiv preprint arXiv:2106.12555}, 2021{\natexlab{a}}.

\bibitem[Dyer et~al.(2021{\natexlab{b}})Dyer, Cannon, and Schmon]{dyer2021deep}
Joel Dyer, Patrick~W Cannon, and Sebastian~M Schmon.
\newblock Deep signature statistics for likelihood-free time-series models.
\newblock In \emph{ICML Workshop on Invertible Neural Networks, Normalizing
  Flows, and Explicit Likelihood Models}, 2021{\natexlab{b}}.

\bibitem[Fearnhead and Prangle(2012)]{Fearnhead2012}
Paul Fearnhead and Dennis Prangle.
\newblock {Constructing summary statistics for approximate Bayesian
  computation: Semi-automatic approximate Bayesian computation}.
\newblock \emph{Journal of the Royal Statistical Society. Series B: Statistical
  Methodology}, 74\penalty0 (3):\penalty0 419--474, 2012.
\newblock ISSN 13697412.
\newblock \doi{10.1111/j.1467-9868.2011.01010.x}.

\bibitem[Franke(2009)]{franke2009applying}
Reiner Franke.
\newblock Applying the method of simulated moments to estimate a small
  agent-based asset pricing model.
\newblock \emph{Journal of Empirical Finance}, 16\penalty0 (5):\penalty0
  804--815, 2009.

\bibitem[Franke and Westerhoff(2012)]{franke2012structural}
Reiner Franke and Frank Westerhoff.
\newblock Structural stochastic volatility in asset pricing dynamics:
  Estimation and model contest.
\newblock \emph{Journal of Economic Dynamics and Control}, 36\penalty0
  (8):\penalty0 1193--1211, 2012.

\bibitem[Gelman et~al.(1996)Gelman, Roberts, Gilks,
  et~al.]{gelman1996efficient}
Andrew Gelman, Gareth~O Roberts, Walter~R Gilks, et~al.
\newblock Efficient metropolis jumping rules.
\newblock \emph{Bayesian statistics}, 5\penalty0 (599-608):\penalty0 42, 1996.

\bibitem[Gon{\c{c}}alves et~al.(2020)Gon{\c{c}}alves, Lueckmann, Deistler,
  Nonnenmacher, {\"O}cal, Bassetto, Chintaluri, Podlaski, Haddad, Vogels,
  et~al.]{gonccalves2020training}
Pedro~J Gon{\c{c}}alves, Jan-Matthis Lueckmann, Michael Deistler, Marcel
  Nonnenmacher, Kaan {\"O}cal, Giacomo Bassetto, Chaitanya Chintaluri,
  William~F Podlaski, Sara~A Haddad, Tim~P Vogels, et~al.
\newblock Training deep neural density estimators to identify mechanistic
  models of neural dynamics.
\newblock \emph{Elife}, 9:\penalty0 e56261, 2020.

\bibitem[Gourieroux et~al.(1993)Gourieroux, Monfort, and
  Renault]{gourieroux1993indirect}
Christian Gourieroux, Alain Monfort, and Eric Renault.
\newblock Indirect inference.
\newblock \emph{Journal of applied econometrics}, 8\penalty0 (S1):\penalty0
  S85--S118, 1993.

\bibitem[Grazzini et~al.(2017)Grazzini, Richiardi, and
  Tsionas]{grazzini2017bayesian}
Jakob Grazzini, Matteo~G Richiardi, and Mike Tsionas.
\newblock Bayesian estimation of agent-based models.
\newblock \emph{Journal of Economic Dynamics and Control}, 77:\penalty0 26--47,
  2017.

\bibitem[Greenberg et~al.(2019)Greenberg, Nonnenmacher, and
  Macke]{Greenberg2019}
David~S. Greenberg, Marcel Nonnenmacher, and Jakob~H. Macke.
\newblock {Automatic posterior transformation for likelihood-free inference}.
\newblock \emph{36th International Conference on Machine Learning, ICML 2019},
  2019-June:\penalty0 4288--4304, 2019.

\bibitem[Gretton et~al.(2006)Gretton, Borgwardt, Rasch, Sch{\"o}lkopf, and
  Smola]{gretton2006kernel}
Arthur Gretton, Karsten Borgwardt, Malte Rasch, Bernhard Sch{\"o}lkopf, and
  Alex Smola.
\newblock A kernel method for the two-sample-problem.
\newblock \emph{Advances in neural information processing systems},
  19:\penalty0 513--520, 2006.

\bibitem[Gretton et~al.(2012)Gretton, Borgwardt, Rasch, Sch{{\"o}}lkopf, and
  Smola]{mmd}
Arthur Gretton, Karsten~M. Borgwardt, Malte~J. Rasch, Bernhard Sch{{\"o}}lkopf,
  and Alexander Smola.
\newblock A kernel two-sample test.
\newblock \emph{Journal of Machine Learning Research}, 13\penalty0
  (25):\penalty0 723--773, 2012.
\newblock URL \url{http://jmlr.org/papers/v13/gretton12a.html}.

\bibitem[Hastings(1970)]{Hastings1970MonteCS}
W.~K. Hastings.
\newblock Monte carlo sampling methods using markov chains and their
  applications.
\newblock \emph{Biometrika}, 57:\penalty0 97--109, 1970.

\bibitem[Hermans et~al.(2020)Hermans, Begy, and Louppe]{pmlr-v119-hermans20a}
Joeri Hermans, Volodimir Begy, and Gilles Louppe.
\newblock Likelihood-free {MCMC} with amortized approximate ratio estimators.
\newblock In Hal~Daumé III and Aarti Singh, editors, \emph{Proceedings of the
  37th International Conference on Machine Learning}, volume 119 of
  \emph{Proceedings of Machine Learning Research}, pages 4239--4248. PMLR,
  13--18 Jul 2020.
\newblock URL \url{https://proceedings.mlr.press/v119/hermans20a.html}.

\bibitem[Ju et~al.(2021)Ju, Heng, and Jacob]{ju2021sequential}
Nianqiao Ju, Jeremy Heng, and Pierre~E Jacob.
\newblock Sequential monte carlo algorithms for agent-based models of disease
  transmission.
\newblock \emph{arXiv preprint arXiv:2101.12156}, 2021.

\bibitem[Kantorovich(1960)]{wass}
L.~V. Kantorovich.
\newblock Mathematical methods of organizing and planning production.
\newblock \emph{Management Science}, 6\penalty0 (4):\penalty0 366--422, 1960.
\newblock ISSN 00251909, 15265501.
\newblock URL \url{http://www.jstor.org/stable/2627082}.

\bibitem[Kidger et~al.(2020)Kidger, Morrill, Foster, and
  Lyons]{kidger2020neural}
Patrick Kidger, James Morrill, James Foster, and Terry Lyons.
\newblock Neural controlled differential equations for irregular time series.
\newblock \emph{arXiv preprint arXiv:2005.08926}, 2020.

\bibitem[Kingma and Ba(2014)]{kingma2014adam}
Diederik~P Kingma and Jimmy Ba.
\newblock Adam: A method for stochastic optimization.
\newblock \emph{arXiv preprint arXiv:1412.6980}, 2014.

\bibitem[Kingma and Dhariwal(2018)]{kingma2018glow}
Diederik~P Kingma and Prafulla Dhariwal.
\newblock Glow: Generative flow with invertible 1x1 convolutions.
\newblock \emph{arXiv preprint arXiv:1807.03039}, 2018.

\bibitem[Koopman(1936)]{koopman1936sufficient}
B.~O. Koopman.
\newblock On distributions admitting a sufficient statistic.
\newblock \emph{Transactions of the American Mathematical Society}, 39\penalty0
  (3):\penalty0 399--409, 1936.
\newblock ISSN 00029947.
\newblock URL \url{http://www.jstor.org/stable/1989758}.

\bibitem[Kukacka and Barunik(2017)]{kukacka2017estimation}
Jiri Kukacka and Jozef Barunik.
\newblock Estimation of financial agent-based models with simulated maximum
  likelihood.
\newblock \emph{Journal of Economic Dynamics and Control}, 85:\penalty0 21--45,
  2017.

\bibitem[Liepe et~al.(2014)Liepe, Kirk, Filippi, Toni, Barnes, and
  Stumpf]{liepe2014framework}
Juliane Liepe, Paul Kirk, Sarah Filippi, Tina Toni, Chris~P Barnes, and
  Michael~PH Stumpf.
\newblock A framework for parameter estimation and model selection from
  experimental data in systems biology using approximate bayesian computation.
\newblock \emph{Nature protocols}, 9\penalty0 (2):\penalty0 439--456, 2014.

\bibitem[Lueckmann et~al.(2017)Lueckmann, Goncalves, Bassetto, {\"O}cal,
  Nonnenmacher, and Macke]{lueckmann2017flexible}
Jan-Matthis Lueckmann, Pedro~J Goncalves, Giacomo Bassetto, Kaan {\"O}cal,
  Marcel Nonnenmacher, and Jakob~H Macke.
\newblock Flexible statistical inference for mechanistic models of neural
  dynamics.
\newblock \emph{Advances in Neural Information Processing Systems}, 30, 2017.

\bibitem[Lueckmann et~al.(2021)Lueckmann, Boelts, Greenberg, Goncalves, and
  Macke]{lueckmann2021benchmarking}
Jan-Matthis Lueckmann, Jan Boelts, David Greenberg, Pedro Goncalves, and Jakob
  Macke.
\newblock Benchmarking simulation-based inference.
\newblock In Arindam Banerjee and Kenji Fukumizu, editors, \emph{Proceedings of
  The 24th International Conference on Artificial Intelligence and Statistics},
  volume 130 of \emph{Proceedings of Machine Learning Research}, pages
  343--351. PMLR, 13--15 Apr 2021.

\bibitem[Lux(2018)]{lux2018estimation}
Thomas Lux.
\newblock Estimation of agent-based models using sequential monte carlo
  methods.
\newblock \emph{Journal of Economic Dynamics and Control}, 91:\penalty0
  391--408, 2018.

\bibitem[Lux(2021)]{lux2021bayesian}
Thomas Lux.
\newblock Bayesian estimation of agent-based models via adaptive particle
  markov chain monte carlo.
\newblock \emph{Computational Economics}, pages 1--27, 2021.

\bibitem[Malleson et~al.(2020)Malleson, Minors, Kieu, Ward, West, and
  Heppenstall]{malleson2020simulating}
Nick Malleson, Kevin Minors, Le-Minh Kieu, Jonathan~A Ward, Andrew West, and
  Alison Heppenstall.
\newblock Simulating crowds in real time with agent-based modelling and a
  particle filter.
\newblock \emph{Journal of Artificial Societies and Social Simulation},
  23\penalty0 (3), 2020.

\bibitem[Noé et~al.(2019)Noé, Olsson, Köhler, and Wu]{noe2019boltz}
Frank Noé, Simon Olsson, Jonas Köhler, and Hao Wu.
\newblock Boltzmann generators: Sampling equilibrium states of many-body
  systems with deep learning.
\newblock \emph{Science}, 365\penalty0 (6457):\penalty0 eaaw1147, 2019.
\newblock \doi{10.1126/science.aaw1147}.
\newblock URL \url{https://www.science.org/doi/abs/10.1126/science.aaw1147}.

\bibitem[Papamakarios and Murray(2016)]{papamakarios2016fast}
George Papamakarios and Iain Murray.
\newblock Fast $\varepsilon$-free inference of simulation models with bayesian
  conditional density estimation.
\newblock In \emph{Advances in neural information processing systems}, pages
  1028--1036, 2016.

\bibitem[Papamakarios et~al.(2017)Papamakarios, Pavlakou, and
  Murray]{papamakarios2017masked}
George Papamakarios, Theo Pavlakou, and Iain Murray.
\newblock Masked autoregressive flow for density estimation.
\newblock In \emph{Proceedings of the 31st International Conference on Neural
  Information Processing Systems}, pages 2335--2344, 2017.

\bibitem[Papamakarios et~al.(2019)Papamakarios, Sterratt, and
  Murray]{papamakarios2019sequential}
George Papamakarios, David Sterratt, and Iain Murray.
\newblock Sequential neural likelihood: Fast likelihood-free inference with
  autoregressive flows.
\newblock In \emph{The 22nd International Conference on Artificial Intelligence
  and Statistics}, pages 837--848. PMLR, 2019.

\bibitem[Park et~al.(2016)Park, Jitkrittum, and Sejdinovic]{Park2016}
Mijung Park, Wittawat Jitkrittum, and Dino Sejdinovic.
\newblock {K2-ABC: Approximate bayesian computation with kernel embeddings}.
\newblock \emph{Proceedings of the 19th International Conference on Artificial
  Intelligence and Statistics, AISTATS 2016}, 41:\penalty0 398--407, 2016.

\bibitem[Paszke et~al.(2019)Paszke, Gross, Massa, Lerer, Bradbury, Chanan,
  Killeen, Lin, Gimelshein, Antiga, et~al.]{paszke2019pytorch}
Adam Paszke, Sam Gross, Francisco Massa, Adam Lerer, James Bradbury, Gregory
  Chanan, Trevor Killeen, Zeming Lin, Natalia Gimelshein, Luca Antiga, et~al.
\newblock Pytorch: An imperative style, high-performance deep learning library.
\newblock \emph{Advances in neural information processing systems},
  32:\penalty0 8026--8037, 2019.

\bibitem[Pitman(1936)]{pitman1936sufficient}
Edwin James~George Pitman.
\newblock Sufficient statistics and intrinsic accuracy.
\newblock In \emph{Mathematical Proceedings of the cambridge Philosophical
  society}, volume~32, pages 567--579. Cambridge University Press, 1936.

\bibitem[Platt(2020)]{PLATT2020103859}
Donovan Platt.
\newblock A comparison of economic agent-based model calibration methods.
\newblock \emph{Journal of Economic Dynamics and Control}, 113:\penalty0
  103859, 2020.
\newblock ISSN 0165-1889.
\newblock \doi{https://doi.org/10.1016/j.jedc.2020.103859}.
\newblock URL
  \url{https://www.sciencedirect.com/science/article/pii/S0165188920300294}.

\bibitem[Platt(2021)]{platt2021bayesian}
Donovan Platt.
\newblock Bayesian estimation of economic simulation models using neural
  networks.
\newblock \emph{Computational Economics}, pages 1--52, 2021.

\bibitem[Price et~al.(2018)Price, Drovandi, Lee, and Nott]{price2018bayesian}
Leah~F Price, Christopher~C Drovandi, Anthony Lee, and David~J Nott.
\newblock Bayesian synthetic likelihood.
\newblock \emph{Journal of Computational and Graphical Statistics}, 27\penalty0
  (1):\penalty0 1--11, 2018.

\bibitem[Pritchard et~al.(1999)Pritchard, Seielstad, Perez-Lezaun, and
  Feldman]{pritchard1999population}
Jonathan~K Pritchard, Mark~T Seielstad, Anna Perez-Lezaun, and Marcus~W
  Feldman.
\newblock Population growth of human {Y} chromosomes: a study of {Y} chromosome
  microsatellites.
\newblock \emph{Molecular biology and evolution}, 16\penalty0 (12):\penalty0
  1791--1798, 1999.

\bibitem[Rezende and Mohamed(2015)]{rezende}
Danilo Rezende and Shakir Mohamed.
\newblock Variational inference with normalizing flows.
\newblock In Francis Bach and David Blei, editors, \emph{Proceedings of the
  32nd International Conference on Machine Learning}, volume~37 of
  \emph{Proceedings of Machine Learning Research}, pages 1530--1538, Lille,
  France, 07--09 Jul 2015. PMLR.
\newblock URL \url{https://proceedings.mlr.press/v37/rezende15.html}.

\bibitem[Roberts et~al.(1997)Roberts, Gelman, and Gilks]{roberts1997}
Gareth~O Roberts, Andrew Gelman, and Walter~R Gilks.
\newblock Weak convergence and optimal scaling of random walk metropolis
  algorithms.
\newblock \emph{The annals of applied probability}, 7\penalty0 (1):\penalty0
  110--120, 1997.

\bibitem[Schmon and Gagnon(2022)]{schmon2021optimal}
Sebastian~M Schmon and Philippe Gagnon.
\newblock Optimal scaling of random walk metropolis algorithms using bayesian
  large-sample asymptotics.
\newblock \emph{Statistics and Computing, forthcoming}, 2022.

\bibitem[Schmon et~al.(2020)Schmon, Cannon, and Knoblauch]{gbiabc}
Sebastian~M Schmon, Patrick~W Cannon, and Jeremias Knoblauch.
\newblock Generalized posteriors in approximate bayesian computation.
\newblock \emph{https://arxiv.org/abs/2011.08644}, 2020.

\bibitem[Schmon et~al.(2021)Schmon, Deligiannidis, Doucet, and
  Pitt]{schmon2021large}
Sebastian~M Schmon, George Deligiannidis, Arnaud Doucet, and Michael~K Pitt.
\newblock Large-sample asymptotics of the pseudo-marginal method.
\newblock \emph{Biometrika}, 108\penalty0 (1):\penalty0 37--51, 2021.

\bibitem[Sherlock et~al.(2015)Sherlock, Thiery, Roberts, and
  Rosenthal]{sherlock2015efficiency}
Chris Sherlock, Alexandre~H Thiery, Gareth~O Roberts, and Jeffrey~S Rosenthal.
\newblock On the efficiency of pseudo-marginal random walk metropolis
  algorithms.
\newblock \emph{The Annals of Statistics}, 43\penalty0 (1):\penalty0 238--275,
  2015.

\bibitem[Sherlock et~al.(2017)Sherlock, Thiery, and Lee]{sherlock2017pseudo}
Chris Sherlock, Alexandre~H Thiery, and Anthony Lee.
\newblock Pseudo-marginal metropolis--hastings sampling using averages of
  unbiased estimators.
\newblock \emph{Biometrika}, 104\penalty0 (3):\penalty0 727--734, 2017.

\bibitem[Shiono(2021)]{shiono2021estimation}
Takashi Shiono.
\newblock Estimation of agent-based models using bayesian deep learning
  approach of bayesflow.
\newblock \emph{Journal of Economic Dynamics and Control}, 125:\penalty0
  104082, 2021.

\bibitem[Silverman(1986)]{silverman1986density}
Bernard~W Silverman.
\newblock \emph{Density Estimation for Statistics and Data Analysis},
  volume~26.
\newblock CRC Press, 1986.

\bibitem[Sunn{\aa}ker et~al.(2013)Sunn{\aa}ker, Busetto, Numminen, Corander,
  Foll, and Dessimoz]{sunnaaker2013approximate}
Mikael Sunn{\aa}ker, Alberto~Giovanni Busetto, Elina Numminen, Jukka Corander,
  Matthieu Foll, and Christophe Dessimoz.
\newblock Approximate bayesian computation.
\newblock \emph{PLoS computational biology}, 9\penalty0 (1):\penalty0 e1002803,
  2013.

\bibitem[Tabak and Turner(2013)]{tabak2013family}
Esteban~G Tabak and Cristina~V Turner.
\newblock A family of nonparametric density estimation algorithms.
\newblock \emph{Communications on Pure and Applied Mathematics}, 66\penalty0
  (2):\penalty0 145--164, 2013.

\bibitem[Tabak and Vanden-Eijnden(2010)]{tabak2010density}
Esteban~G Tabak and Eric Vanden-Eijnden.
\newblock Density estimation by dual ascent of the log-likelihood.
\newblock \emph{Communications in Mathematical Sciences}, 8\penalty0
  (1):\penalty0 217--233, 2010.

\bibitem[Talts et~al.(2020)Talts, Betancourt, Simpson, Vehtari, and
  Gelman]{talts2020validating}
Sean Talts, Michael Betancourt, Daniel Simpson, Aki Vehtari, and Andrew Gelman.
\newblock Validating bayesian inference algorithms with simulation-based
  calibration, 2020.

\bibitem[Tavar{\'e} et~al.(1997)Tavar{\'e}, Balding, Griffiths, and
  Donnelly]{tavare1997inferring}
Simon Tavar{\'e}, David~J Balding, Robert~C Griffiths, and Peter Donnelly.
\newblock Inferring coalescence times from dna sequence data.
\newblock \emph{Genetics}, 145\penalty0 (2):\penalty0 505--518, 1997.

\bibitem[Tejero-Cantero et~al.(2020)Tejero-Cantero, Boelts, Deistler,
  Lueckmann, Durkan, Gonçalves, Greenberg, and Macke]{tejero-cantero2020sbi}
Alvaro Tejero-Cantero, Jan Boelts, Michael Deistler, Jan-Matthis Lueckmann,
  Conor Durkan, Pedro~J. Gonçalves, David~S. Greenberg, and Jakob~H. Macke.
\newblock sbi: A toolkit for simulation-based inference.
\newblock \emph{Journal of Open Source Software}, 5\penalty0 (52):\penalty0
  2505, 2020.
\newblock \doi{10.21105/joss.02505}.
\newblock URL \url{https://doi.org/10.21105/joss.02505}.

\bibitem[Thomas et~al.(2021)Thomas, Dutta, Corander, Kaski, and
  Gutmann]{thomas2021lfire}
Owen Thomas, Ritabrata Dutta, Jukka Corander, Samuel Kaski, and Michael~U.
  Gutmann.
\newblock {Likelihood-Free Inference by Ratio Estimation}.
\newblock \emph{Bayesian Analysis}, pages 1 -- 31, 2021.
\newblock \doi{10.1214/20-BA1238}.
\newblock URL \url{https://doi.org/10.1214/20-BA1238}.

\bibitem[Toni et~al.(2009)Toni, Welch, Strelkowa, Ipsen, and
  Stumpf]{toni2009approximate}
Tina Toni, David Welch, Natalja Strelkowa, Andreas Ipsen, and Michael~PH
  Stumpf.
\newblock Approximate bayesian computation scheme for parameter inference and
  model selection in dynamical systems.
\newblock \emph{Journal of the Royal Society Interface}, 6\penalty0
  (31):\penalty0 187--202, 2009.

\bibitem[Wilkinson(2013)]{wilkinson2013approximate}
Richard~David Wilkinson.
\newblock Approximate bayesian computation (abc) gives exact results under the
  assumption of model error.
\newblock \emph{Statistical applications in genetics and molecular biology},
  12\penalty0 (2):\penalty0 129--141, 2013.

\bibitem[Wiqvist et~al.(2019)Wiqvist, Mattei, Picchini, and
  Frellsen]{wiqvist2019partially}
Samuel Wiqvist, Pierre-Alexandre Mattei, Umberto Picchini, and Jes Frellsen.
\newblock {Partially exchangeable networks and architectures for learning
  summary statistics in approximate Bayesian computation}.
\newblock In \emph{International Conference on Machine Learning}, pages
  6798--6807. PMLR, 2019.

\bibitem[Wong et~al.(2018)Wong, Jiang, Wu, and Zheng]{Wong_2018}
Wing Wong, Bai Jiang, Tung-yu Wu, and Charles Zheng.
\newblock Learning summary statistic for approximate bayesian computation via
  deep neural network.
\newblock \emph{Statistica Sinica}, 2018.
\newblock ISSN 1017-0405.
\newblock \doi{10.5705/ss.202015.0340}.
\newblock URL \url{http://dx.doi.org/10.5705/ss.202015.0340}.

\bibitem[Wood(2010)]{wood2010statistical}
Simon~N Wood.
\newblock Statistical inference for noisy nonlinear ecological dynamic systems.
\newblock \emph{Nature}, 466\penalty0 (7310):\penalty0 1102--1104, 2010.

\bibitem[Zaheer et~al.(2017)Zaheer, Kottur, Ravanbakhsh, Poczos, Salakhutdinov,
  and Smola]{NIPS2017_f22e4747}
Manzil Zaheer, Satwik Kottur, Siamak Ravanbakhsh, Barnabas Poczos, Russ~R
  Salakhutdinov, and Alexander~J Smola.
\newblock Deep sets.
\newblock In I.~Guyon, U.~V. Luxburg, S.~Bengio, H.~Wallach, R.~Fergus,
  S.~Vishwanathan, and R.~Garnett, editors, \emph{Advances in Neural
  Information Processing Systems}, volume~30. Curran Associates, Inc., 2017.
\newblock URL
  \url{https://proceedings.neurips.cc/paper/2017/file/f22e4747da1aa27e363d86d40ff442fe-Paper.pdf}.

\end{thebibliography}

\clearpage

\appendix

\section{Posterior sampling}

\subsection{Sampling with Metropolis-Hastings}\label{app:mh}

The Metropolis-Hastings algorithm is a classical algorithm for generating samples from some density $p(\bth)$. Starting from $\bth_0$ and given a proposal distribution $q( \cdot \mid \bth)$ which proposes successive values in the chain, it entails repeating the following steps: at each step $t \geq 1$,

\begin{enumerate}
    \item propose $\bth \sim q(\cdot \mid \bth_t)$;
    \item accept $\bth$ (i.e. set $\bth_{t+1} := \bth$) with probability
    \begin{equation*}
        \alpha = \min\Bigg\{ 1 , \frac{p(\bth)q(\bth_t \mid \bth)}{p(\bth_t)q( \bth\mid \bth_t)} \Bigg\},
    \end{equation*}
    else set $\bth_{t+1} = \bth_t$.
\end{enumerate}
Since such a procedure generates correlated samples from the posterior, it is customary to \emph{thin} the samples by retaining every $n$th sample for some integer $n \geq 1$ chosen as required.

To sample from the posteriors obtained via \km{} and \gls{nre} in Section \ref{sec:tractable}, we use \gls{mh} with a normal proposal distribution $q(\cdot \mid \bth) = \mathcal{N}(\bth; \ell^2\Sigma)$, and perform a trial run of 50,000 steps using an isotropic Gaussian to estimate the covariance matrix $\Sigma$ of $q$ before a further 100,000 steps are run. The use of such pilot runs is long standing-practice for random walk \gls{mh} algorithms and can be motivated theoretically, for example, using Bayesian asymptotics \citep[][]{schmon2021optimal}. In addition, we set $\ell = 2/\sqrt{d}$, where $d$ is the parameter dimension following the guidelines of \citet{gelman1996efficient, roberts1997, schmon2021optimal}. 
All chains are initialised at the parameter value which generated the observation. We thin the resultant chains by retaining every $100$th value, resulting in 1,000 approximately uncorrelated samples from the respective posteriors.

\subsection{Sampling with sampling-importance-resampling}\label{app:sir}
\glsresetall

\Gls{sir} is an approach to obtaining samples from a target distribution $f(\bth)$ given samples from a different distribution $g(\bth)$ which proceeds as follows:
\begin{enumerate}
    \item generate samples $\lbrace{\bth_i\rbrace}_{i=1}^{N} \overset{iid}{\sim} g(\bth)$;
    \item compute weights $w_i = f(\bth_i)/g(\bth_i)$ for all $i$;
    \item resample $\bth_i$ with probability proportional to $w_i$.
\end{enumerate}
In particular, this may be used in density ratio estimation to approximate the posterior by setting $g(\bth) := p(\bth)$, $f(\bth) := p(\bth \mid \by)$, and using that the density ratio estimator estimates the ratio
\begin{equation}
    \frac{f(\bth)}{g(\bth)} = \frac{p(\bth \mid \by)}{p(\bth)} =  \frac{p(\by \mid \bth)}{p(\by)}.
\end{equation}

\section{Neural network architectures and training}\label{app:nn}

For each neural network method, the $z$-scores of all variables are taken prior to passing them into the networks. For all neural posterior estimation tasks, we use a masked autoregressive flow \citep{papamakarios2017masked} with 5 flow transforms, each with 2 blocks and 50 hidden features; for all neural density ratio estimation tasks, we use a residual network with two layers of size 50. To train the network weights, we use Adam \citep{kingma2014adam}, along with a training batch size of 50 and learning rate of $5\times 10^{-4}$. We furthermore reserve 10\% of the data for validation, and stop training when the validation error does not improve over 20 epochs to avoid overfitting. Throughout, we use the \texttt{sbi} python package \citep{tejero-cantero2020sbi}.

\end{document}